\documentclass[10pt]{article}
\usepackage{amsmath,amssymb,graphicx}
\usepackage{hyperref}
\usepackage[T1]{fontenc}
\usepackage{baskervald}
\usepackage[font=small,labelfont=bf]{caption}
\usepackage{longtable}
\usepackage{verbatim}
\usepackage{amsfonts}
\usepackage{cite}

\DeclareMathAlphabet\mathbfcal{OMS}{cmsy}{b}{n}

\usepackage[usenames,dvipsnames]{xcolor}
\definecolor{darkerblue}{rgb}{0.2,0.2,0.5}

\definecolor{seagreen}{rgb}{0.180392,0.545098,0.341176}

\usepackage{afterpage}

\newcommand{\bear}{\begin{array}}
\newcommand{\ear}{\end{array}}

\newcommand{\be}{\begin{equation}}
\newcommand{\ee}{\end{equation}}
\newcommand{\ba}{\begin{align}}
\newcommand{\ea}{\end{align}}
\newcommand{\beq}{\begin{eqnarray}}
\newcommand{\eeq}{\end{eqnarray}}
\newcommand{\beqa}{\begin{eqnarray}}
\newcommand{\eeqa}{\end{eqnarray}}

\def\OMIT#1{{}}
\newcommand{\lsim}{\mathrel{\rlap{\lower4pt\hbox{\hskip1pt$\sim$}}
    \raise1pt\hbox{$<$}}}         
\newcommand{\gsim}{\mathrel{\rlap{\lower4pt\hbox{\hskip1pt$\sim$}}
    \raise1pt\hbox{$>$}}}         
\def\simlt{\stackrel{<}{{}_\sim}}
\def\simgt{\stackrel{>}{{}_\sim}}

\usepackage{tikz}
\usetikzlibrary{trees}
\usetikzlibrary{decorations.pathmorphing}
\usetikzlibrary{decorations.markings}
\usetikzlibrary{positioning}
\tikzset{
    photon/.style={decorate, decoration={snake,amplitude=3pt,segment length=8pt}, draw=black},
    wino/.style={draw=redwine},    
    fermion/.style={draw=black, postaction={decorate},
        decoration={markings,mark=at position .55 with {\arrow[draw=black,scale=2,#1]{>}}}},
    scalar/.style={draw=black, dashed,postaction={decorate},
        decoration={markings,mark=at position .55 with {\arrow[draw=black,scale=2,#1]{>}}}},
    gluon/.style={decorate, draw=black,
        decoration={coil,amplitude=3pt, segment length=4pt}},
    graviton/.style={decorate, draw=black,
        decoration={zigzag,amplitude=3pt, segment length=4pt}}
}
\tikzstyle{blob}=[circle,
                              minimum size=0.01,
                              draw=black!80,
                              fill=black!80]   
\tikzstyle{redblob}=[circle,
                              thick,
                              minimum size=0.4cm,
                              draw=red!80,
                              fill=red!60] 

\usepackage{titlesec}

\title{\bf SUSY's Ladder:\\ 
Reframing Sequestering at Large Volume}

\author{Matthew Reece$^a$ and Wei Xue$^b$\\
{\small \color{gray} \texttt{mreece@physics.harvard.edu, weixue@mit.edu}}\\
{$^a$ \em Department of Physics, Harvard University, Cambridge, MA 02138, USA}\\
{$^b$ \em Center for Theoretical Physics, Massachusetts Institute of Technology,}\\{\em Cambridge, MA 02139, USA}}

\begin{document}
\maketitle

\vspace{-9.5cm}
\begin{flushright}
{\footnotesize MIT-CTP-4749} 
\end{flushright}
\vspace{8.1cm}

\begin{abstract}
Theories with approximate no-scale structure, such as the Large Volume Scenario, have a distinctive hierarchy of multiple mass scales in between TeV gaugino masses and the Planck scale, which we call SUSY's Ladder. This is a particular realization of Split Supersymmetry in which the same small parameter suppresses gaugino masses relative to scalar soft masses, scalar soft masses relative to 
the gravitino mass, and the UV cutoff or string scale relative to the Planck scale.
This scenario has many phenomenologically interesting properties, 
and can avoid dangers including the gravitino problem, flavor problems, and the moduli-induced LSP problem that plague other supersymmetric theories.
We study SUSY's Ladder using a superspace formalism that makes the mysterious cancelations in previous computations manifest. This opens the possibility of a consistent effective field theory understanding of the phenomenology of these scenarios, based on power-counting in the small ratio of string to Planck scales. We also show that four-dimensional theories with approximate 
no-scale structure enforced by a single volume modulus arise only from two special higher-dimensional theories: five-dimensional supergravity and 
ten-dimensional type IIB supergravity. This gives a phenomenological argument in favor of ten dimensional ultraviolet physics which is different from standard arguments based on the consistency of superstring theory.
\end{abstract}
\renewcommand*\contentsname{}
\tableofcontents

\section{Introduction}
\label{sec:intro}

An appealing feature of supersymmetry is the automatic presence of WIMP dark matter candidates in the form of gauginos or higgsinos. Often these are discussed in a thermal scenario assuming the universe is populated by a hot plasma of MSSM fields. The gravitino, as an extremely weakly coupled particle that will not be in thermal equilibrium, can complicate or even ruin this appealing story~\cite{Pagels:1981ke,Weinberg:1982zq,Khlopov:1984pf}. This motivates the question: how much can we reasonably expect the gravitino to be decoupled from gauginos? In this paper, we will explore this question, largely from a bottom-up effective field theory viewpoint. Our arguments suggest that the simplest, most robust way to decouple the gravitino is through no-scale structure \cite{Cremmer:1983bf,Ellis:1983sf,Ellis:1984bm} enforced by the volume modulus of either a single extra dimension or of six extra dimensions arising from Type IIB supergravity in ten dimensions. The argument that singles out 10d Type IIB is completely independent of any consideration related to string theory. We will explore how no-scale structure can suppress various SUSY-breaking effects, showing that a variety of initially surprising results previously derived in the Large Volume Scenario of string theory \cite{Balasubramanian:2005zx,Conlon:2005ki} are readily understood by working in superspace in the conformal compensator formalism and choosing not to work in Einstein frame. We view our results as a step toward bridging the gap between bottom-up field theory and top-down string constructions. Phenomenologically, we are led to an interesting version of split SUSY with scalar superpartners at the PeV scale.

\begin{figure}[h]
\begin{center}
\begin{tikzpicture}[line width=1.5 pt]
\node at (15.2,0.5) {$m_{3/2}~[{\rm TeV}]$};
\draw[postaction={decorate},decoration={markings,mark=at position 1.0 with {\arrow[draw=black]{>}}}] (0,0)--(16,0);
\draw[dashed] (4,0.05)--(4,-3);
\node at (4,0.35) {$m_{\chi^0_1}$};
\draw[dashed] (8,0.05)--(8,-3);
\node at (8,0.35) {$100~{\rm TeV}$};
\draw[dashed] (12,0.05)--(12,-3);
\node at (12,0.35) {$10^4~{\rm TeV}$};
\node[text centered] at (2,-0.5) {I:~$m_{3/2} < m_{\chi^0_1}$};
\node[text centered] at (2,-1.0) {\bf Low-scale};
\node[text width=100,align=left,anchor=north] at (2,-1.5) {Gravitino overclosure. Late-time thermal inflation?};
\node[text centered] at (6,-0.5) {II:~$T_{3/2} < T_{\rm BBN}$};
\node[text centered] at (6,-1.0) {\bf BBN concerns};
\node[text width=100,align=left,anchor=north] at (6,-1.5) {Must avoid producing gravitinos, thermally and during reheating.};
\node[text centered] at (10,-0.5) {III:~$T_{\rm BBN} < T_{3/2} < T_{\rm FO}$};
\node[text centered] at (10,-1.0) {\bf Dark matter concerns};
\node[text width=100,align=left,anchor=north] at (10,-1.5) {Nonthermal histories. Moduli dilution. Data increases tension.};
\node[text centered] at (14,-0.5) {IV:~$T_{\rm FO} < T_{3/2}$};
\node[text width=100,align=left,anchor=north] at (14,-1.5) {Conventional thermal history. Calls for strong sequestering.};
\node[text centered] at (14,-1.0) {\bf Safe region};
\end{tikzpicture}
\end{center}
\caption{Regimes of gravitino mass with distinct cosmology. Conventional thermal relic SUSY WIMPs are most straightforwardly achieved in the rightmost region, Regime IV: $m_{3/2} \simgt 10^4~{\rm TeV}$. In Regimes II and III, $m_{\chi^0_1} < m_{3/2} \simlt 10^4~{\rm TeV}$, gravitino production must be suppressed to have a conventional thermal relic WIMP.} \label{fig:gravitinoregimes}
\end{figure}
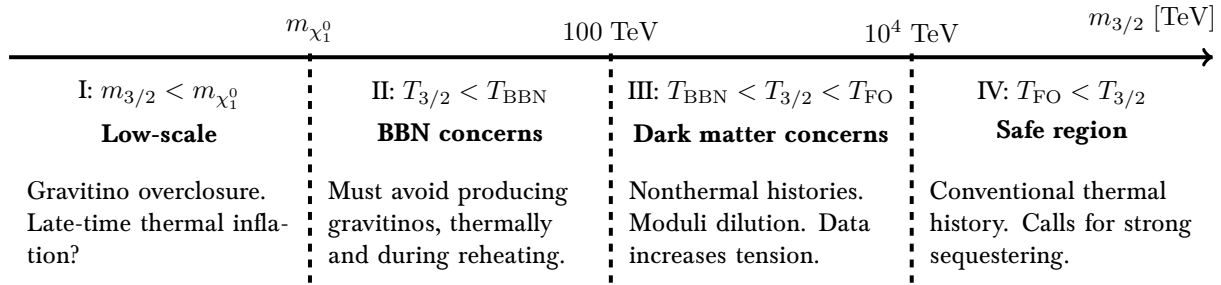

We are motivated to study sequestered theories due to the prospect for a clean solution to cosmological gravitino~\cite{Pagels:1981ke,Weinberg:1982zq,Khlopov:1984pf} and moduli~\cite{Ellis:1986zt,Coughlan:1983ci,deCarlos:1993jw} problems. We will focus our discussion on the gravitino, since it necessarily exists in any supersymmetric completion of the Standard Model. (We expect that moduli fields also exist in any theory of quantum gravity, but this is less obvious.) In the MSSM, the gravitino decay width assuming ample phase space for all decays is $\Gamma_{3/2} = \frac{193}{384\pi} m_{3/2}^3/M_{\rm Pl}^2$~\cite{Moroi:1995fs,Nakamura:2006uc}. This implies that gravitinos in a radiation-dominated universe decay when the temperature is about
\beq
T_{3/2} \approx 10~{\rm MeV} \left(m_{3/2}/100~{\rm TeV}\right)^{3/2}.
\eeq
This separates the cosmology of supersymmetric theories into several different regimes depending on the gravitino mass and its relation to the lightest Standard Model superpartner mass, as illustrated in Figure~\ref{fig:gravitinoregimes}. Gravitinos can be produced in the early universe through thermal scattering \cite{Ellis:1984eq,Pradler:2006hh,Rychkov:2007uq}, freeze-in \cite{Cheung:2011nn}, decays of superpartners after freeze-out \cite{Feng:2003xh,Feng:2004mt} and inflaton or moduli decays \cite{Kawasaki:2006gs,Dine:2006ii, Kawasaki:2006hm, Endo:2006zj}, so there is a potentially problematic population of these particles in the early universe. The first case is when the gravitino is lighter than $m_{\chi^0_1}$. In this case, Standard Model superpartner masses must lie below about 10 TeV~\cite{Hall:2013uga}. Light gravitinos pose a variety of cosmological problems that are compounded if moduli fields are also light~\cite{Moroi:1993mb,deGouvea:1997tn,Choi:1998dw}. We will not discuss this low-scale SUSY breaking scenario further in this paper. Regime II, in which gravitinos decay after Big Bang Nucleosynthesis has already begun, is particularly dangerous \cite{Kawasaki:2004qu,Kawasaki:2008qe,Kawasaki:2015yya,Cyburt:2015mya}. In Regime III, decays of the gravitino are unconstrained by BBN but invalidate the standard thermal WIMP freeze-out calculations, generally leading to a larger dark matter abundance that risks overclosing the universe \cite{Moroi:2013sla}, especially in light of indirect detection constraints \cite{Fan:2014gxa}. Again, adding moduli to the scenario tends to increase the dark matter abundance \cite{Moroi:1999zb,Acharya:2008bk,Acharya:2009zt,Kane:2015jia} and indirect detection makes this ``moduli-induced LSP problem'' more severe \cite{Cohen:2013ama,Fan:2013faa,Blinov:2014nla,Cheung:2014hya}. It is only in Regime IV that gravitinos are, from a cosmological viewpoint, a non-issue: they decay above the thermal WIMP freeze-out temperature, and so conventional WIMP relic abundance calculations are directly applicable. The threshold gravitino mass for which this applies is
\beq
m_{3/2} \simgt 10^4~{\rm TeV}.
\eeq
Cosmological moduli (or modulinos, saxions, or other long-lived weakly-coupled particles) present similar issues and generally make the gravitino problem worse. But the gravitino problem already gives us a clear justification for seeking a scenario in which some superpartners have masses at around the 100 GeV to TeV scale and constitute WIMP candidates while the gravitino mass is at least four orders of magnitude larger.

Scenarios in which the gravitino mass is parametrically larger than the Standard Model superpartner masses are known as ``sequestered'' theories. In their original incarnations, they led to gravitino masses one loop factor larger than SM superpartner masses, allowing anomaly-mediated terms to dominate \cite{Randall:1998uk, Luty:2001jh}. This is an interesting scenario, but it does not immediately solve the gravitino problem, because it leads to Regime III: $m_{3/2} \sim 10^2~{\rm to}~10^3~{\rm TeV}$ for weak-to-TeV scale WIMPs. We need a more powerful version of sequestering. One way to see the generic difficulty of sequestering is the following: we work in superspace with the conformal compensator field $\bf \Phi$ \cite{Cremmer:1982en}. Throughout the paper we will follow a convention of writing superfields (like $\bf Q$) in boldface, using the same notation without boldface  (e.g.~$Q$) for the scalar component of a chiral multiplet, a subscripted $\psi$ for the fermion component (e.g.~$\psi_Q$), and a subscripted $F$ for the auxiliary field (e.g.~$F_Q$). In superspace, kinetic terms for chiral multiplets take the form
\beq
\int d^4 \theta {\bf \Phi}^\dagger {\bf \Phi} {\bf Q}^\dagger {\bf Q}.
\eeq
In the simplest theories one finds that ${\bf \Phi} = 1 + m_{3/2} \theta^2$ and so this term immediately produces a scalar mass term $m_{3/2}^2 Q^\dagger Q$. Similarly, one-loop corrections dependent on $\log(\Lambda {\bf \Phi}/\mu_{\rm RG})$ 
produce gaugino masses $m_\lambda \sim (\alpha/\pi) m_{3/2}$ \cite{Giudice:1998xp,Randall:1998uk}. This suggests that what we need is a mechanism to suppress the $\theta^2$ component of $\bf \Phi$.

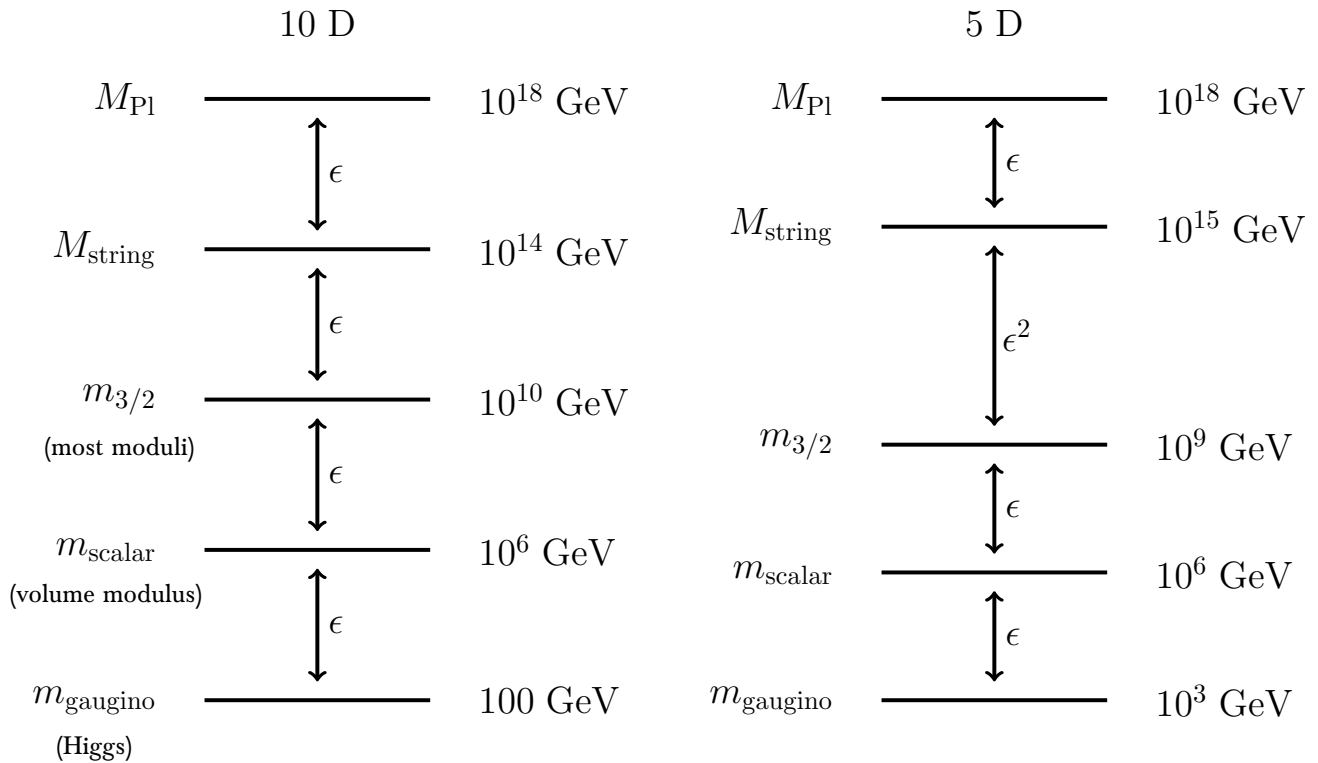
\begin{figure}[h]
\begin{center}
\begin{tikzpicture}[line width=1.5 pt]
\node[above] at (1.5, 10.7 ) {\Large\bf $10 ~ \mathrm{D}$ };
\draw (0,10)--(3,10); \node[left] at (-0.5,10) {\Large $M_{\rm Pl}$}; \node[right] at (3.5,10) {\Large $10^{18}~{\rm GeV}$};
\draw[<->] (1.5,9.75)--(1.5,8.25);
\node at (1.75,9) {\Large $\epsilon$};
\draw (0,8)--(3,8); \node[left] at (-0.5,8) {\Large $M_{\rm string}$}; \node[right] at (3.5,8) {\Large $10^{14}~{\rm GeV}$};
\draw[<->] (1.5,7.75)--(1.5,6.25);
\node at (1.75,7) {\Large $\epsilon$};
\draw (0,6)--(3,6); \node[left] at (-0.5,6) (a) {\Large $m_{3/2}$}; \node[right] at (3.5,6) {\Large $10^{10}~{\rm GeV}$};
\node[below =0.1mm of a] {(most moduli)};
\draw[<->] (1.5,5.75)--(1.5,4.25);
\node at (1.75,5) {\Large $\epsilon$};
\draw (0,4)--(3,4); \node[left] at (-0.5,4) (b) {\Large $m_{\rm scalar}$}; \node[right] at (3.5,4) {\Large $10^{6}~{\rm GeV}$};
\node[below =0.1mm of b] {(volume modulus)};
\draw[<->] (1.5,3.75)--(1.5,2.25);
\node at (1.75,3) {\Large $\epsilon$};
\draw (0,2)--(3,2); \node[left] at (-0.5,2) (c) {\Large $m_{\rm gaugino}$}; \node[right] at (3.5,2) {\Large $100~{\rm GeV}$};
\node[below =0.1mm of c] {(Higgs)};
\node[above] at (10.5, 10.7 ) {\Large\bf $5 ~ \mathrm{D}$ };
\draw (9,10)--(12,10); \node[left] at (8.5,10) {\Large $M_{\rm Pl}$}; \node[right] at (12.5,10) {\Large $10^{18}~{\rm GeV}$};
\draw[<->] (10.5,9.75)--(10.5,8.55);
\node at (10.75,9.15) {\Large $\epsilon$};
\draw (9,8.3)--(12,8.3); \node[left] at (8.5,8.3) {\Large $M_{\rm string}$}; \node[right] at (12.5,8.3) {\Large $10^{15}~{\rm GeV}$};
\draw[<->] (10.5,8.05)--(10.5,5.65);
\node at (10.8,6.85) {\Large $\epsilon^2$};
\draw (9,5.4)--(12,5.4); \node[left] at (8.5,5.4) (a) {\Large $m_{3/2}$}; \node[right] at (12.5,5.4) {\Large $10^{9}~{\rm GeV}$};
\draw[<->] (10.5,5.15)--(10.5,3.95);
\node at (10.75,4.55) {\Large $\epsilon$};
\draw (9,3.7)--(12,3.7); \node[left] at (8.5,3.7) (b) {\Large $m_{\rm scalar}$}; \node[right] at (12.5,3.7) {\Large $10^{6}~{\rm GeV}$};
\draw[<->] (10.5,3.45)--(10.5,2.25);
\node at (10.75,2.85) {\Large $\epsilon$};
\draw (9,2)--(12,2); \node[left] at (8.5,2) (c) {\Large $m_{\rm gaugino}$}; \node[right] at (12.5,2) {\Large $10^3~{\rm GeV}$};
\end{tikzpicture}
\end{center}
\caption{``SUSY's Ladder'': Predicted spectrum in 10d~(left) and 5d~(right). The Higgs is tuned to be light, but other fields are at their natural values for the case ${\epsilon} \sim 10^4$ in 10d and ${\epsilon} \sim 10^3$ in 5d. We will focus most of our attention on the 10d scenario, which was previously obtained as the ``local scenario'' of \cite{Aparicio:2014wxa} (following earlier work in \cite{Blumenhagen:2009gk}). The logic leading to this picture is explained in \S\ref{sec:hierarchies}.} \label{fig:spectrum}
\label{fig:ladder}
\end{figure}

It has been appreciated for many years that no-scale structure can suppress soft masses \cite{Binetruy:1985ap,Binetruy:1987xj} and even anomaly-mediation effects \cite{Luty:1999cz,Bagger:1999rd,Gaillard:2000fk}. This is readily understood in superspace \cite{Luty:1999cz,Luty:2002hj,ArkaniHamed:2004fb}: the simplest realization of no-scale structure is
\beq
\int d^4 \theta {\bf \Phi}^\dagger {\bf \Phi} ({\bf T} + {\bf T}^\dagger), \label{eq:linearterm}
\eeq
essentially just introducing a Lagrange multiplier to force $F_\Phi = 0$. This eliminates the most dangerous effects that are naively proportional to $m_{3/2}$. Of course, the real challenge is to ask what subleading terms involve ${\bf T}$, which in realistic examples is a physical field and not just a Lagrange multiplier. One must ensure that the leading linear term has a dominant influence on the physics. Theories with more complicated kinetic functions involving multiple fields can also lead to $F_\Phi = 0$, so we refer to the case with a linear term as ``single-field no-scale structure.'' Because it is the simplest version of no-scale structure, we expect that it is the most robust (i.e.~that the corrections can be controlled in the most straightforward way). In this paper we will limit our attention to the single-field version of no-scale structure.

Our goal is to explore sequestered scenarios driven by no-scale structure, largely taking a bottom-up viewpoint that complements work done in the string phenomenology context. In section \ref{sec:purenoscale}, we review the basic mechanism by which no-scale structure suppresses soft terms. We also look at how the scalar field $T$ controlling no-scale structure can have highly SUSY-violating decays. In the pure no-scale limit, it turns out that the modulus $T$ can decay to higgs fields through the $\mu$-term, but does not decay to 
the superpartner, higgsinos. Then the production of dark matter and the process of thermalization will be different from the standard 
WIMP scenario. In section \ref{sec:dimred}, we ask when single-field no-scale structure can arise from compactifying extra dimensions. We focus on the case when the single field corresponds to an isotropic rescaling of the compactified directions, and argue that no-scale structure arises in two special cases: compactification of five-dimensional supergravity or of ten-dimensional Type IIB supergravity. Importantly, this is an argument based purely on the desire to find a linear term like (\ref{eq:linearterm}) in the action, and proceeds entirely at the level of effective field theory. As such, this argument is completely independent of the sort of consistency arguments that single out ten dimensions in the string theory context. It is compelling that bottom-up, phenomenologically motivated reasoning highlights 10d Type IIB as a special theory, without UV input. 
In section \ref{sec:hierarchies}, we study the one-loop corrections from a heavy gravitino, and find that it naturally introduces hierarchies between gravitino, scalar and gaugino when there is a cutoff $\Lambda$ well below the Planck scale $M_{\rm Pl}$. Scalar mass squares and gaugino masses are 
suppressed by the quadratic divergences from one-loop, $\frac{\Lambda^2}{M_{\rm Pl}^2}$.
Meanwhile, the gravitino mass is related to the superpotential and no-scale structure. All these things lead to
SUSY's Ladder as shown in fig.~\ref{fig:ladder}: a stratified spectrum with several large gaps all set by a single parametrically small number $\epsilon$. (Such a spectrum was first obtained in the string phenomenology context in \cite{Blumenhagen:2009gk}.) Following the line of effective field theory, in section \ref{sec:deviations} we show possible ways to deviate from pure no-scale structure, such as by introducing other moduli; and we compute the F-terms of 
moduli and study the moduli stabilization. Having the F-terms for the moduli fields, we study the soft SUSY breaking 
terms in more detail in section \ref{sec:susybreaking}. We find two important results. The first is that it is actually possible to realize the SUSY's Ladder spectrum of hierarchies that we estimated on generic grounds. The second is that working in superspace with the conformal compensator, rather than going to Einstein frame and working in components as is typically done in the literature, makes it easy to simply read off the parametric size of all the SUSY-breaking effects without any apparent cancelations. Finally, we conclude in section \ref{sec:discussion} with a brief sketch of two directions that will require future work. The first future direction is the phenomenology resulting from SUSY's Ladder. We argue that this scenario works very well for achieving a realistic Higgs mass of 125 GeV, starting from approximately universal scalar masses at a high scale which lead to $\tan \beta \approx 2$ and PeV scalars. The scenario also is appealing from the standpoint of flavor physics and offers novel cosmological possibilities for obtaining the correct WIMP dark matter abundance, in accord with our original motivation of avoiding gravitino problems. The second future direction is a more thorough study of effective field theory and loop corrections to the scenario. Ideally, we would have a supersymmetric formalism on the same footing with our tree-level use of superspace and the conformal compensator. As a step in this direction, we compute the effective K\"ahler potential and show that the quadratically divergent terms have the right parametric dependence on $\epsilon$ to lead to small corrections to our tree-level results. This is an encouraging indication of the consistency of the whole picture.

\section{Pure no-scale: phenomenological results}
\label{sec:purenoscale}

In this section, we will review some of the features of no-scale structure in the simplest possible context of pure, single-field, no-scale structure: that in which a chiral superfield $\bf T$ appears {\em only} linearly in the kinetic function $\bf \Omega$. We show that this structure can lead to suppression of soft SUSY-breaking effects including the conformal compensator $F$-term (which in turn suppresses anomaly mediated effects). We also point out that this structure allows highly supersymmetry-violating decays of the field $T$: the scalar modulus can have a nonzero decay width to the scalar component of a chiral supermultiplet but a zero decay width to the fermionic component. These results are known in the literature, but we give a streamlined presentation of them in superspace that eliminates the apparent cancelations that have appeared in some of the previous derivations (though not all~\cite{Higaki:2012ar}). In later sections we will look at how additional terms can lead to violations of no-scale structure while still preserving the dominant phenomenological features at leading order.

\subsection{Single-field no-scale structure}

We begin with a toy Lagrangian capturing the idea of pure, single-field, no-scale structure. It has a chiral superfield ${\bf T}$ appearing linearly in the kinetic function $\bf \Omega$ and absent from the superpotential:
\be
{\cal L}_{\rm 0} = -3 \int d^4 \theta \, {\bf \Phi}^\dagger {\bf \Phi} M_*^2 ({\bf T} + {\bf T}^\dagger) + \left(\int d^2 \theta \, {\bf \Phi}^3 W_0 + {\rm h.c.}\right).
\ee
The normalization of the graviton kinetic term requires that $\left<\Phi^\dagger \Phi (T+T^\dagger)\right> M_*^2 = M_{\rm Pl}^2$. The terms in ${\cal L}_0$ involving the $F$ components $F_\Phi = \left. {\bf \Phi} \right|_{\theta^2}$ and $F_T = \left. {\bf T} \right|_{\theta^2}$ are
\be
\left. {\cal L}_0 \right|_F = -3 |F_\Phi|^2 M_*^2 (T + T^\dagger) - 3 M_*^2 \left(\Phi^\dagger F_\Phi F_T^\dagger + F_\Phi \psi_\Phi^\dagger \psi_T^\dagger + {\rm h.c.}\right) + \left(3 \Phi^2 F_\Phi W_0 + {\rm h.c.}\right).
\ee
From this we can read off immediately two equations of motion:
\begin{align}
\frac{\delta}{\delta F_T^\dagger}: \quad & F_\Phi = 0. \label{eq:Fphipuresol}\\
\frac{\delta}{\delta F_\Phi^\dagger}: \quad & F_T =  \frac{(\Phi^\dagger)^2}{\Phi} \frac{W_0^\dagger}{M_*^2} - \frac{\psi_T \psi_\Phi}{\Phi}\label{eq:FTpuresol}
\end{align}
The fact that $F_\phi = 0$ removes many effects of SUSY-breaking, including anomaly mediated effects.

\subsection{Kinetic unmixing and the Cheung/D'Eramo/Thaler gauge}

The function ${\bf \Omega} = {\bf T} + {\bf T}^\dagger$ has the unusual feature of being linear in the field ${\bf T}$. In particular, if we ignore the conformal compensator field $\bf \Phi$, it would appear that ${\bf T}$ has no kinetic term. However, $\bf \Omega$ is not the K\"ahler potential but is related to it via ${\bf \Omega} = \exp(-{\bf K}/(3 M_{\rm Pl}^2))$. The familiar component formalism for supergravity tells us that derivatives of $K = -3 M_{\rm Pl}^2 \log(T + T^\dagger)$ determine the kinetic term of the scalar component of $\bf T$. One way to see this from the superspace viewpoint is to recall that, although we have not explicitly written the vielbein multiplet in our action, it is present and wherever we see $\int d^4 \theta\, {\bf \Phi}^\dagger {\bf \Phi}$ in superspace there will be a term $\propto \sqrt{-g} {\cal R}(g)$ in components, with ${\cal R}$ the Ricci scalar. As a result, there is a kinetic mixing of the scalar $T$ with the graviton, which we can remove by going to Einstein frame.

If $T$ were a real field $T = T_0 + \delta T$, we would have the problem of removing the quadratic mixing from the action
\be
{\cal L}_{\rm mixed} = -\frac{M_{\rm Pl}^2}{2} \int d^4 x \sqrt{-g} {\cal R} \left(1 + \delta T/T_0\right).
\ee
This is accomplished with the rescaling ${\tilde g}_{\mu \nu} = g_{\mu \nu} \xi(T)$ where $\xi(T)$ is any function with Taylor expansion $\xi(T) = 1  + \delta T/T_0 + {\cal O}(\delta T)^2$. This produces an independent healthy (positive) kinetic term for the scalar,
\be
\int d^4 x \sqrt{-g}\, 3 M_{\rm Pl}^2 \partial_\mu (\delta T) \partial^\mu (\delta T)/T_0^2.
\ee
In principle, we could proceed in this manner to disentangle kinetic mixing effects.

It would be more convenient, however, to maintain manifest supersymmetry throughout this process rather than writing down kinetic terms in components. Fortunately, an elegant procedure for achieving this has been recently explained by Cheung, D'Eramo, and Thaler \cite{Cheung:2011jp,Cheung:2011jq} (improving on an earlier related idea \cite{Kugo:1982mr}). We will refer to it as the CDT gauge. This gauge choice essentially builds the supersymmetric Weyl transformation into the choice of conformal compensator field, removing all linear terms in chiral superfields from the product ${\bf \Phi}^\dagger {\bf \Phi} {\bf \Omega}$. The procedure is to fix the gauge so that
\begin{align}
{\bf \Phi} &= e^{{\bf Z}/3}(1 + f_\Phi \theta^2) \\
{\bf Z} &= \frac{1}{M_{\rm Pl}^2} \left[\left<K/2 - i M_{\rm Pl}^2\, {\rm arg} W\right> + \left<K_i\right>({\bf X^i} - \left<X^i\right>)\right],
\end{align}
and as CDT show the $F_\phi^\dagger$ equation of motion sets $\left<f_\Phi\right> = m_{3/2}$. We have altered their notation to emphasize that $f_\Phi$ is {\em not} the $\theta^2$ component of the chiral superfield $\bf \Phi$, because ${\bf Z}$ itself can have a $\theta^2$ component. By our previous observation, for pure no-scale structure the $F_T^\dagger$ equation of motion enforces $F_\Phi \equiv \left. \Phi\right|_{\theta^2} = 0$, despite the fact that $f_\Phi \neq 0$.

In the case of single-field no-scale structure, we have
\begin{align}
{\bf Z} &= -\frac{3}{2} \log(\left<T + T^\dagger\right>) - i\, {\rm arg}\, W_0 - \frac{3}{\left<T + T^\dagger\right>} ({\bf T} - \left<T\right>), \\
{\bf \Phi} &= \frac{1}{\left<T + T^\dagger\right>^{1/2}} e^{- ({\bf T} - \left<T\right>)/\left<T + T^\dagger\right> - i({\rm arg}\, W_0)/3}\left(1 + f_\Phi \theta^2\right).
\end{align}
With this gauge choice we should take $M_* = M_{\rm Pl}$ to normalize the graviton kinetic term appropriately, and expanding out the exponential we find a term
\be
-3 M_{\rm Pl}^2 {\bf \Phi}^\dagger {\bf \Phi} ({\bf T} + {\bf T}^\dagger) \supset 3 M_{\rm Pl}^2 \frac{{\bf T}^\dagger {\bf T}}{\left<T + T^\dagger\right>^2}.
\ee
Thus we define a canonically normalized modulus superfield
\be
{\bf T}^c = \sqrt{3} M_{\rm Pl} \frac{{\bf T} - \left<T\right>}{\left<T + T^\dagger\right>}.
\ee
Our earlier results (\ref{eq:Fphipuresol}, \ref{eq:FTpuresol}) have the consequences that
\begin{align}
f_\Phi &= \frac{\left. {\bf T}^c \right|_{\theta^2}}{\sqrt{3} M_{\rm Pl}}, \\
\left<F_T\right> &= \frac{\left<\Phi^\dagger\right>^2}{\left<\Phi\right>} \frac{W_0^\dagger}{M_{\rm Pl}^2} = \frac{\left|W_0\right|}{\left<T + T^\dagger\right>^{1/2} M_{\rm Pl}^2}, \\
m_{3/2} &= \left<f_\Phi\right> = \frac{\left<F_T\right>}{\left<T + T^\dagger\right>} = \frac{\left|W_0\right|}{\left<T + T^\dagger\right>^{3/2} M_{\rm Pl}^2},
\end{align}
in accord with the familiar result $m_{3/2} = e^{\left<K/(2 M_{\rm Pl}^2)\right>} \left<W\right>/M_{\rm Pl}^2$. From now on, we will set ${\rm arg}\, W_0 = 0$ for simplicity. In terms of the canonically normalized field, we have
\be
{\bf \Phi} = \frac{1}{\left<T + T^\dagger\right>^{1/2}} e^{-{\bf T}^c/(\sqrt{3} M_{\rm Pl})} \left(1 + \frac{\left. {\bf T}^c \right|_{\theta^2}}{\sqrt{3} M_{\rm Pl}} \theta^2 \right). \label{eq:PhiCDT}
\ee
Notice that CDT gauge maintains the fact that the net $\theta^2$ component of $\bf \Phi$ is zero while ensuring that $\bf \Phi$ has linear pieces in all {\em other} components of the chiral multiplet ${\bf T}$ in order to disentangle the kinetic mixing arising from the linear terms in $\bf \Omega$. One way to say this is that at linear order, ${\bf \Phi}$ contains a modified modulus field with no $F$-component, which we denote ${\bf \tilde T}^c$:
\begin{align}
{\bf \Phi} &= \frac{1}{\left<T + T^\dagger\right>^{1/2}} \left(1 - \frac{1}{\sqrt{3} M_{\rm Pl}} {\bf \tilde T}^c + \ldots\right), \\
{\bf \tilde T}^c &\equiv {\bf T}^c - \left. {\bf T}^c \right|_{\theta^2} \theta^2.
\end{align}

\subsection{Sequestered chiral superfields}
\label{subsec:sequesteredchiral}

Suppose now that we add to the pure no-scale Lagrangian one or more chiral superfields that do not directly couple to $\bf T$, but only to the conformal compensator. In order to discuss both soft masses and $\mu$-terms, we add two fields ${\bf Q}$ and ${\bf \bar Q}$: 
\be
{\cal L} = {\cal L}_0 + \int d^4 \theta\, {\bf \Phi}^\dagger {\bf \Phi} \left[{\bf Q}^\dagger {\bf Q} + {\bf \bar Q}^\dagger {\bf \bar Q} + \left(z {\bf \bar Q} {\bf Q} + {\rm h.c.}\right)\right].
\ee
This corresponds to a K\"ahler potential with the sequestered form $K = - 3 M_{\rm Pl}^2 \log(T + T^\dagger - \frac{1}{3 M_{\rm Pl}^2} Q^\dagger Q - \ldots)$. Our first observation is that in the pure no-scale limit, the fact that $F_\Phi = 0$ implies that these terms lead to no soft mass, $\mu$ term, or $b$ term. However, this does not mean that SUSY-breaking effects are completely absent here. They are present in the couplings of the modulus ${\bf T}$ to the fields ${\bf Q}$ and ${\bf \bar Q}$ that are induced by the Weyl transformation that removes the kinetic mixing of the modulus and gravity. In CDT gauge, these couplings are manifest through the ${\bf T}$ dependence of ${\bf \Phi}$.

Using the expression (\ref{eq:PhiCDT}) for the conformal compensator, we see that the new terms in our Lagrangian correspond to
\be
\int d^4 \theta \frac{1}{\left<T + T^\dagger\right>} e^{-({\bf T}^c + {\bf T}^{c\dagger})/(\sqrt{3} M_{\rm Pl})} \left(1 + \frac{\left. {\bf T}^c \right|_{\theta^2} \theta^2 + \left. {\bf T}^{c\dagger} \right|_{\theta^{\dagger 2}} \theta^{\dagger 2}}{\sqrt{3} M_{\rm Pl}}  \right) \left[{\bf Q}^\dagger {\bf Q} + {\bf \bar Q}^\dagger {\bf \bar Q} + \left(z {\bf \bar Q} {\bf Q} + {\rm h.c.}\right)\right].
\ee
We define canonically normalized quark fields, ${\bf Q}^c = \frac{1}{\left<T +T^\dagger\right>^{1/2}} {\bf Q}$. In terms of this field, one of the terms in our Lagrangian is
\be
- \int d^4 \theta \frac{{\bf \tilde T}^c + {\bf \tilde T}^{c\dagger}}{\sqrt{3} M_{\rm Pl}} {\bf Q}^{c\dagger} {\bf Q}^c.
\ee
This gives rise to couplings
\be
\frac{1}{\sqrt{3} M_{\rm Pl}} \left[ (T^c Q^c)^\dagger \Box Q^c + (T^c Q^c) \Box Q^{c\dagger}\right],
\ee
so that for on-shell decays we see that the amplitude for $T^c$ decays to squarks is proportional to the squark mass-squared. A similar statement holds for the decays to fermions arising from the fermion kinetic term. Similar observations were already made in \cite{Endo:2006xg}. The utility of the conformal compensator formalism for studying questions like this one has been remarked upon in \cite{Endo:2007sz,Higaki:2012ar}.

In the case of holomorphic ($\mu$-term like) couplings, we get a similar structure
\be
-\int d^4 \theta \frac{{\bf \tilde T}^c + {\bf \tilde T}^{c\dagger}}{\sqrt{3} M_{\rm Pl}} \left(z {\bf {\bar Q}}^c {\bf Q}^c + {\rm h.c.}\right),
\ee
giving rise to 
\be
\frac{1}{\sqrt{3} M_{\rm Pl}} \left[ z (\Box T^c) {\bar Q}^{c\dagger} Q^{c\dagger} + {\rm h.c.}\right],
\ee
so that the decay amplitude of the modulus to scalars through such a coupling is proportional to the {\em modulus} mass. Importantly, however, there is no corresponding coupling to fermions, because such a coupling would arise from the $\theta^2$ component that is absent in ${\bf \tilde T}^c$.

\subsection{Implications for the moduli-induced dark matter problem}

Let us pause here to emphasize the importance of the result we have just derived. In the pure no-scale limit, the modulus ${\bf T}$ can decay to scalar fields through a $\mu$-term like coupling {\em without} having a comparable decay rate to the fermionic superpartners of those scalars. This is a dramatic violation of supersymmetry. For instance, in the case of a chiral superfield ${\bf X}$ with a superpotential $\frac{1}{2} M_X {\bf X}^2$ and a coupling $\int d^4 \theta ({\bf X} + {\bf X}^\dagger) {\bf \bar Q} {\bf Q}$, the derivation of the decay to scalars $X \to {\bar Q} Q$ will proceed exactly as we have derived above. The decay to fermions will be dramatically different, however, because of the coupling $F_X^\dagger \psi_{\bar Q} \psi_Q$ and the equation of motion $F_X^\dagger = m_X X$. This ensures that the two decay rates are equal, $\Gamma(X \to {\bar Q} Q) = \Gamma(X \to \psi_{\bar Q} \psi_Q) \propto m_X^3$. This result follows from a supersymmetric Ward identity \cite{Bose:2013fqa} and ensures that for ``typical'' moduli, even after taking supersymmetry breaking into account, decays to particles and their superpartners will have comparable rates \cite{Kaplan:2006vm, Nakamura:2006uc}.

On the contrary, decays of a no-scale modulus---whose couplings inherently violate supersymmetry---can be maximally far from democratic. The modulus can decay to a particle while having, to leading approximation, zero decay rate to that particle's superpartner. This has been previously pointed out in \cite{Cicoli:2012aq, Higaki:2012ar}. It is potentially of great importance for the phenomenology of supersymmetric dark matter. In the standard nonthermal modulus decay scenario \cite{Moroi:1999zb}, the modulus has a significant decay rate to $R$-parity odd particles which can then annihilate significantly (e.g.~for wino dark matter) to lead to a sufficiently small final abundance. A no-scale modulus, on the other hand, has the potential to decay overwhelmingly to $R$-even particles, so that $R$-odd particles are produced either with a small branching ratio or through scattering of the $R$-even particles. If the branching fraction is sufficiently small, the final abundance could arise from freeze-in of such scattering processes \cite{Chung:1998rq}. The unique possibilities of no-scale structure for the nonthermal dark matter scenario have only begun to be explored \cite{Allahverdi:2013noa, Allahverdi:2014ppa}.

\subsection{Integrating out heavy moduli}
\label{subsec:integrateout}

As a first step toward investigating the robustness of the above results, suppose there is another modulus which is stabilized with a large mass, e.g.
\be
{\cal L} = -3 \int d^4\theta\, {\bf \Phi}^\dagger {\bf \Phi} M_{\rm Pl}^2 ({\bf T} + {\bf T}^\dagger)({\bf S} + {\bf S}^\dagger)^{1/3} + \left[\int d^2 \theta \, {\bf \Phi}^3 \left(W_0 + \frac{1}{2} W_S ({\bf S} - S_0)^2\right) + {\rm h.c.}\right].
\ee
It is clear that the superpotential wants to set $S = S_0$, and that to the extent that $W_S$ is very large, $S$ should then decouple and leave the no-scale structure intact. Let us be more explicit about this. In particular, it could be the case that $S$ controls the strength of a gauge interaction:
\be
{\cal L}_{\rm gauge} = \int d^2 \theta {\bf S} \mathbfcal{W}^\alpha \mathbfcal{W}_\alpha + {\rm h.c.}
\ee
In this case, the gaugino mass depends on $F_S$, which is nearly zero to the extent that $W_S$ is very large, but in general will be nonzero due to the interaction between $T$ and $S$. One common way to approach this problem is simply to solve for the VEVs and $F$-terms of all fields, including $\bf S$. But a nicer way is to supersymmetrically integrate out the chiral superfield $\bf S$, so that in the low-energy effective theory we can make the replacement ${\bf S} \to S_0 + f(\bf T)$ for some function $f$, so that rather than explicitly solving for $F_S$ we relate the gaugino mass to $F_T$.

To do this, write ${\bf S} = S_0 + {\bf s}$ and expand the Lagrangian in $\bf s$:
\begin{align}
{\cal L} = & -3 \int d^4 \theta \, {\bf \Phi}^\dagger {\bf \Phi} M_{\rm Pl}^2 ({\bf T} + {\bf T}^\dagger)(S_0 + S_0^\dagger)^{1/3}\left(1 + \frac{{\bf s}+{\bf s}^\dagger}{3(S_0+S_0^\dagger)} - \frac{({\bf s}+{\bf s}^\dagger)^2}{9(S_0+S_0^\dagger)^2}\right) \nonumber \\
& + \left[\int d^2 \theta \, \left\{{\bf \Phi}^3 \left(W_0 + \frac{1}{2} W_S {\bf s}^2\right) + (S_0 + {\bf s}) \mathbfcal{W}^\alpha \mathbfcal{W}_\alpha\right\} + {\rm h.c.}\right].
\end{align}
From this we find the equation of motion by varying with respect to $\bf s$:
\be
W_S \Phi^3 {\bf s} + \mathbfcal{W}^\alpha \mathbfcal{W}_{\alpha} = 3 \int d^2 {\bar \theta}{\bf \Phi}^\dagger {\bf \Phi} M_{\rm Pl}^2 \frac{{\bf T} + {\bf T}^\dagger}{3(S_0 + S_0^\dagger)^{2/3}}\left(1 - \frac{2({\bf s} + {\bf s}^\dagger)}{9(S_0+S_0^\dagger)}\right).
\ee
Substituting this solution back into the original Lagrangian, we find a variety of terms, including higher-dimension gauge interactions like $\int d^2\theta\, \frac{1}{{\bf \Phi}^3 W_S}(\mathbfcal{W}^\alpha \mathbfcal{W}_\alpha)^2$. But our immediate interest is mostly in the gaugino mass, which will arise from the term:
\be
\int d^4 \theta M_{\rm Pl}^2 \frac{{\bf T} + {\bf T}^\dagger}{(S_0+S_0^\dagger)^{2/3}} \left(\frac{{\bf \Phi}^\dagger}{{\bf \Phi}^2 W_S} \mathbfcal{W}^\alpha \mathbfcal{W}_\alpha + {\rm h.c.}\right).
\ee
However, there is {\em no} gaugino mass associated with this term! The gaugino mass comes in only at higher order, after we add no-scale breaking terms that will lead to $\left<F_\Phi\right> \neq 0$.

Notice that, due to the additional factor in front of $M_{\rm Pl}^2$, in this context correctly normalizing the Einstein-Hilbert term will lead to an additional factor in the VEV of the compensator field,
\be
\left<\Phi\right> = \frac{1}{\left<T + T^\dagger\right>^{1/2} \left<S + S^\dagger\right>^{1/6}}.
\ee

\section{Dimensional reduction: why no-scale?}
\label{sec:dimred}

In this section, we will review some aspects of compactified gravitational theories and choices of frame in which to study them. We argue that the single-field no-scale structure that we studied in \S\ref{sec:purenoscale} is a good approximation only for two very special classes of four-dimensional theories: those arising from compactification of a five-dimensional supergravity theory and those arising from compactification of a ten-dimensional supergravity theory that contains four-form gauge fields (which singles out Type IIB supergravity). Interestingly, this argument for considering an underlying ten dimensional theory is unrelated to the standard arguments for the critical dimension of superstrings.

\subsection{Three frames for Kaluza-Klein reductions}
\label{subsec:kk}

Consider a compactification from $D = d+n$ spacetime dimensions down to $d$ spacetime dimensions~\cite{Bailin:1987jd}. In general, the lower-dimensional theory contains a variety of fields arising from the higher-dimensional metric, including the lower-dimensional metric, scalar and vector modes, and their heavier Kaluza-Klein cousins. For the purposes of this discussion, we will only consider fluctuations of the overall volume of the higher dimensions. In other words, we will break the $D$-dimensional coordinates into $d$-dimensional coordinates $x^\mu$ and $n$-dimensional coordinates $y^m$ and consider a scalar metric perturbation $L(x)$ (the ``volume modulus'') of the form:
\beq
ds^2 = g_{\mu \nu}(x) dx^\mu dx^\nu + L(x)^2 h_{lm}(y) dy^l dy^m.
\eeq
Notice that $L(x)$ corresponds to an isotropic change of the length scale of the internal dimensions (and that it is dimensionless). The Einstein-Hilbert term in the $d$-dimensional action resulting from this ansatz will not be canonically normalized, as it will contain a factor of the $n$-dimensional volume proportional to $L(x)^n$ that leads to kinetic mixing between the scalar field $L$ and the graviton. A Weyl rescaling of the metric disentangles the kinetic term of the metric and that of the scalar $L$, taking us to Einstein frame.

Assuming that the metric on the internal space, $h_{lm}$, is Ricci-flat (as it is e.g.~for a torus or a Calabi-Yau manifold), we find that the higher-dimensional Einstein-Hilbert term reduces to
\beq
{\cal L}_{\rm EH} = -\frac{1}{16\pi G_{D}} \int d^d x \sqrt{-g} V_n L(x)^n \left({\cal R}_d + n(n-1) g^{\mu \nu} \partial_\mu (\log L) \partial_\nu (\log L)\right).
\eeq
Here ${\cal R}_d$ is the Ricci scalar associated with the $d$-dimensional metric $g_{\mu \nu}$, $G_{D}$ is the higher-dimensional Newton's constant, $V_n = \int d^n y \sqrt{\left|h\right|}$, and the $d$-dimensional Newton's constant is 
\beq
G_d = \frac{G_{D}}{V_n \left<L\right>^{n}}.
\eeq
We will also find it useful to write $M_{\rm Pl} = (8 \pi G_d)^{-1/(d-2)}$ for the lower-dimensional Planck scale and ${\tilde M}_{\rm Pl} = (8 \pi G_D)^{-1/(D-2)}$ for the higher-dimensional one, so that $M_{\rm Pl}^{d-2} = V_n \left<L\right>^{n} {\tilde M}_{\rm Pl}^{D-2}$.
Notice that if $n = 1$, the field $L$ has no kinetic term aside from its kinetic mixing with the graviton. This results in an approximate no-scale structure arising when compactifying 5d supergravity theories to four dimensions~\cite{Scherk:1979zr,Kobayashi:2000ak,Chacko:2000fn,Kaplan:2001cg,Luty:2002hj,ArkaniHamed:2004fb}. We will return to this point in more detail below.

We can eliminate the kinetic mixing between the volume modulus and the graviton by doing a Weyl rescaling of the metric, $g_{\mu \nu} \to (L(x)/\left<L\right>)^{-2n/(d-2)} g_{\mu \nu}$. Dropping total derivatives, this results in the action
\beq
{\cal L}_{\rm Einstein} = -\frac{1}{16\pi G_d} \int d^d x \sqrt{-g} \left({\cal R}_d - \frac{n(n+d-2)}{d-2} \partial_\mu (\log L) \partial^\mu (\log L)\right).
\eeq
This is the standard Einstein frame Lagrangian which is most commonly used for calculations; a simple rescaling by a constant now suffices to canonically normalize the field $L$.

For our purposes it is interesting to consider a different Weyl rescaling, namely one which {\em removes} the kinetic term for $L(x)$, so that it propagates only by mixing with the graviton. In some sense this is the frame that is maximally far from Einstein frame, but it is a very useful one; we can refer to it as the keinstein frame.\footnote{Leaving no stone unturned in the pursuit of bad physics puns.} For the special case $n = 1$, we already saw that the simple compactification ansatz led directly to such a Lagrangian. For other values of $n$, one can show that an appropriate Weyl rescaling leads to
\beq
{\cal L}_{\rm kein} = -\frac{1}{16 \pi G_d} \int d^d x \sqrt{-g} {\cal R}_d L^{\sqrt{\frac{n(n+d-2)}{d-1}}} = -\frac{1}{16 \pi G_d} \int d^d x \sqrt{-g} {\cal R}_d L^{\alpha},
\eeq
where $\alpha \equiv \sqrt{n(n+d-2)/(d-1)}$. Now, $L$ sets a characteristic length scale in the internal dimensions: the internal volume, in particular, goes as $L^n$. So cases where $\alpha$ is an integer can be particularly interesting, as the factor multiplying ${\cal R}_d$ may have an interpretation as the volume of an $\alpha$-dimensional structure within the $n$-dimensional internal space. There are precisely two such integer cases, for compactifications of higher dimensional theories down to $d = 4$ dimensions:
\begin{itemize}
\item $n = 1$: In this case, $\alpha = 1$, and we reproduce our earlier result for compactifications from five dimensions down to four.
\item $n = 6$: In this case, $\alpha = 4$, and we find that the action scales as the fourth power of the internal length scale or, equivalently, the volume to the two-thirds power. This corresponds to compactifications from ten dimensions down to four, as in superstring theory~\cite{Scherk:1974ca,Gliozzi:1976jf,Witten:1985xb,Polchinski:1998rr}.
\end{itemize}
More generally, away from four dimensions, the case $n=1$ always yields $\alpha = 1$. If we restrict to $D = d+n \leq 11$, to have the potential of finding supergravity theories, there are precisely three other special cases: $d = 3, n = 8, \alpha = 6$;  $d = 6, n = 5, \alpha = 3$; and $d = 7, n = 3, \alpha = 2$. Intriguingly, all of the special cases originate in theories in 10 or 11 dimensions. There are many other integer solutions at larger values of $D$, beginning at $D=16$, but it seems unlikely that there is interesting physics associated with them. This is as far as it is useful to go in the nonsupersymmetric setting. Now let us turn to supersymmetry to see why this frame is of particular interest.

\subsection{Chiral superfields and no-scale structure}

The above discussion at first glance seems to be a purely academic exercise: we are free to translate between frames as we like, and there is no deep physical principle telling us that $L(x)^\alpha$ is a more useful parametrization of the underlying field for any particular value of $\alpha$ or that keinstein frame is preferred to any other. This changes in the case of supersymmetry. We would like to find approximate pure no-scale structure, i.e.~a Lagrangian that is well-approximated by ${\bf \Phi}^\dagger {\bf \Phi} ({\bf T} + {\bf T}^\dagger)$. Because the Ricci scalar ${\cal R}$ lives in the supergravity multiplet, this means that we would like to ask: when does Kaluza-Klein reduction of a higher-dimensional {\em supergravity} theory lead to 
\be
L^\alpha = T + T^\dagger,
\ee
so that the action is linear in the {\em real part of a chiral superfield}? We could approach this question by systematically studying the different higher-dimensional supergravity theories and repeating the exercise above in the SUSY setting. Instead, let us give a faster argument that we expect will identify all of the relevant cases.

If the Lagrangian depends on ${\bf T}$ (mostly) through the combination ${\bf T} + {\bf T}^\dagger$, the imaginary part of the complex scalar $T$ has an (approximate) shift symmetry. In other words, given the parametrization
\beq
T = \tau + i b + \sqrt{2} \theta \psi_T + \theta^2 F_T + {\rm derivatives},
\eeq
our action is (nearly) independent of $b$. Such a symmetry in the compactified theory should originate with some sort of symmetry in the parent higher-dimensional theory. In fact, such shift symmetries are very familiar consequences of supergravity theories containing $p$-form gauge fields $B_{\mu_1 \ldots \mu_p}$. After we compactify, we obtain a scalar field $b = \int_\Sigma B$ for any $p$-cycle $\Sigma$ in the compactification manifold. The field $b$ is compact, i.e.~it has a gauged discrete shift symmetry originating from large gauge transformations of $B$ in the original theory.

Because $\tau \sim L^\alpha$ and $L$ is a mode of the higher-dimensional graviton, the gauge field we are looking for should be {\em part of} the supergravity multiplet. Because $b$ is obtained by integrating the gauge field over a $p$-dimensional submanifold, we expect $\tau$ to be related to the volume of that submanifold. Hence, we are looking for supergravity theories containing $p$-form gauge fields with 
\beq
p = \alpha = \sqrt{\frac{n(n+d-2)}{d-1}}.
\eeq
We have already identified the cases of integer $\alpha$, so it remains to check whether a $p$-form gauge field of appropriate rank exists in the corresponding supergravity theories. For four dimensional theories we find precisely two examples in which isotropic rescaling of the internal dimensions can be associated with a chiral superfield with no-scale structure:
\begin{itemize}
\item $n = 1, p = 1$: This is the case of a five-dimensional supergravity theory, in which a one-form $A_M(x,y)$ gives rise to a four-dimensional axion $a(x) = \oint dy A_5(x,y)$ that is paired with the radion in a 4d chiral superfield.
\item $n = 6, p = 4$: This is the case of 10-dimensional Type IIB supergravity, which contains a four-form $C_{MNPQ}$ that gives rise to a four-dimensional axion when integrated over a four-dimensional cycle in the geometry. The volume of this four-cycle is the real part of the K\"ahler modulus $T$, and the six-dimensional internal volume scales as ${\cal V} \propto ({\rm Re}~T)^{3/2}$.
\end{itemize}
The arguments given here clarify why four-dimensional no-scale supergravity has been studied in the two contexts of five-dimensional supergravity compactified on a circle and ten-dimensional Type IIB superstring theory compactified on Calabi-Yau manifolds~\cite{Giddings:2001yu}. The absence of no-scale structure in ten-dimensional Type I supergravity (including heterotic strings) and Type IIA supergravity, for instance, is due to the lack of a four-form in the supergravity multiplet, so that the chiral superfields corresponding to moduli organize themselves into multiplets which always have an independent kinetic term that, independent of one's choice of frame, is not acquired by mixing with the graviton. (We explain the relationship of past claims of no-scale structure in heterotic string theory, which apply only after certain fields are integrated out, to our statements here in Appendix \ref{app:heterotic}.) Away from four dimensional compactifications, the special case of $d = 3, n = 8, p = 6$ has been observed to lead to no-scale structure in compactifications of M-theory on Calabi-Yau four-folds \cite{Berg:2002es}. It is unclear whether any physical significance can be attached to the other special cases $d = 6, n = 5, p = 3$ and $d = 7, n = 3, p = 2$.

It is noteworthy that no-scale structure, which has interesting phenomenological properties, provides an argument for considering the compactification of Type IIB ten-dimensional supergravity theory. Notice that superstring theory and the usual worldsheet consistency arguments selecting ten dimensions played no role in our discussion. On the other hand, we expect that the existence of branes as supergravity solitons and the existence of the fundamental string as a soliton of the $D$-brane worldvolume theory \cite{Callan:1997kz} make it essentially inevitable that the only UV complete version of Type IIB supergravity is the Type IIB superstring. Attempting to decouple the gravitino from cosmology while maintaining gauginos near the weak scale has led us (with a few plausible assumptions along the way) to a bottom-up argument for Type IIB superstrings as a UV completion.

We emphasize that our arguments are restricted to {\em single-field} no-scale structure: the simple case in which $F_T$ approximately acts as a Lagrange multiplier setting $F_\Phi = 0$. This includes cases where additional moduli are present but are either massive and decouple or play a subleading role, which we will consider in \S\ref{sec:deviations}.\footnote{An anonymous referee has pointed out to us that \S4.1 of \cite{Conlon:2007dw} gives an argument that even in IIB theories with many K\"ahler moduli, precisely the isotropic rescaling field we have considered is the one that breaks SUSY in the no-scale limit (i.e.~its superpartner is the goldstino). This clarifies the generality of our approach.} One interesting possible example is obtaining the 5d no-scale scenario from 11d heterotic M-theory in the limit where the moduli controlling a Calabi-Yau factor in the compactification are much heavier than the modulus controlling the radius of the M-theory interval. We comment on this case in Appendix \ref{app:heterotic}. We have nothing to say about cases where $F_\Phi = 0$ is enforced by multiple fields, although these cases are also referred to as no-scale structure in the literature. Furthermore, we have assumed a specific geometric origin in the rescaling of internal dimensions.  The string duality web suggests that there may be related theories, for instance in the Type IIA context \cite{Palti:2008mg}, that realize similar physics but where the chiral superfield enforcing no-scale structure no longer has such a simple geometric origin.

\section{Hierarchies in sequestered theories}
\label{sec:hierarchies}

\subsection{Limiting hierarchies with quadratic divergences}
\label{sec:gravitinoloop}

Let us step back and review our logic so far. We would like to consider a theory where the gravitino is relatively heavy compared to other superpartners; in Regime IV of Fig.~\ref{fig:gravitinoregimes}, we required $m_{3/2} \simgt 10^4~{\rm TeV}$, while a neutralino dark matter candidate will have $m_{\chi^0_1} \simlt 1~{\rm TeV}$. How can we achieve a hierarchy of at least four orders of magnitude between gravitino and gaugino masses? We saw in \S\ref{sec:purenoscale} that no-scale structure can suppress tree-level gaugino masses (as well as other SUSY-breaking effects) and in  \S\ref{sec:dimred} that no-scale structure can arise at tree-level from dimensional reduction of certain special supergravity theories in higher dimensions. But we are seeking a quantum theory, not a classical one. Will such a theory be consistent beyond tree-level? In this subsection we will present an argument that it can, precisely in a theory with a UV cutoff well below the Planck scale---a situation that obtains automatically when we compactify large extra dimensions!

The potential problem with consistent no-scale structure arises from divergent radiative corrections. Although soft supersymmetry breaking removes quadratic divergences arising from renormalizable interactions in the Standard Model, nonrenormalizable supersymmetric theories do contain power divergences~\cite{Bagger:1993ji,Choi:1994xg,Bagger:1995ay,Nilles:1997me,Choi:1997de,Brignole:2000kg,Gaillard:2005cw}. These divergences have the potential to destabilize large hierarchies and invalidate extreme realizations of sequestering. A simple dimensional analysis argument sheds light on the largest gaps in the spectrum that we can expect to be consistent with the existence of such divergences. We will simply compute the expected size of loop corrections to scalar and gaugino soft masses arising from loops of gravitons. Similar estimates have been performed in refs.~\cite{Gaillard:1982rm,Barbieri:1983af,Luty:2002ff,Lee:2013aia} and are closely related to spurion arguments in refs.~\cite{ArkaniHamed:2004fb,ArkaniHamed:2004yi}.

\begin{figure}[!h]
\begin{center}
\resizebox{\textwidth}{!}{
\begin{tikzpicture}
\node at (-1,-0.25) {$\phi$};
\node at (-0.2,0.75) {$h_{\mu \nu}$};
\draw[dashed] (-1,0)--(3,0);
\draw[graviton] (1,0.76) circle (0.739);
\node[blob] at (1,0) {};
\begin{scope}[shift={(4,0)}]
\node at (0,-0.25) {$\phi$};
\node at (2.0,1.5) {$h_{\mu \nu}$};
\draw[dashed] (0,0)--(4,0);
\draw[graviton] (3,0.1) arc(0:180:1);
\node[blob] at (1,0) {};
\node[blob] at (3,0) {};
\end{scope}
\begin{scope}[shift={(9,0)}]
\node at (0,-0.25) {$\phi$};
\node at (2,-0.3) {$\psi$};
\node at (2,1.5) {$\psi_\mu$};
\draw[dashed] (0,0)--(1,0);
\draw[fermion] (1,0)--(3,0);
\draw[graviton] (3,0) arc(0:180:1);
\draw[fermion] (3,0) arc(0:180:1);
\draw[dashed] (3,0)--(4,0);
\node[blob] at (1,0) {};
\node[blob] at (3,0) {};
\end{scope}
\begin{scope}[shift={(15,0)}]
\node at (-1,-0.25) {$\phi$};
\node at (-0.2,0.75) {$\psi_\mu$};
\draw[dashed] (-1,0)--(3,0);
\draw[graviton] (1,0.76) circle (0.739);
\draw[fermion] (1,0.76) circle (0.739);
\node[blob] at (1,0) {};
\end{scope}
\end{tikzpicture} 
}
\end{center}
\caption{Gravitational loop corrections to scalar masses. Similar diagrams exist at two loops attaching graviton lines to one-loop diagrams involving renormalizable couplings of $\phi$. These pictures are an oversimplification: the blobs must include enough structure to make the lack of a shift symmetry on $\phi$ manifest.} 
\label{fig:scalargravcorr}
\end{figure}
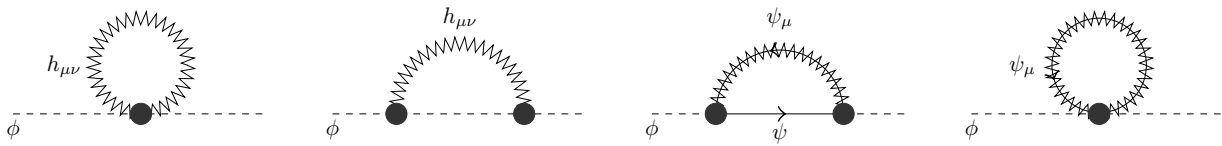

In the case of scalar masses, we consider diagrams such as those shown in Fig.~\ref{fig:scalargravcorr}. At first glance, dimensional analysis tells us that the graviton loop diagrams would be quartically divergent in a nonsupersymmetric theory, but adding the gravitino loop ameliorates the divergence. The divergent part of the result scales as
\beq
\delta_{\rm grav} m_\phi^2 \approx \frac{m_{3/2}^2}{16\pi^2 M_{\rm Pl}^2} \left(a \Lambda^2 + b m_{3/2}^2 \log\Lambda\right).
\eeq
This estimate is generically correct, although the numbers $a$ and $b$ may be somewhat less than order one. The reason is that we do not expect perturbative corrections to violate shift symmetries (e.g.~for an axion field which can get a mass only via instantons). As a result, some aspect of the calculation must be able to tell the difference between a generic scalar field $\phi$ (like the Higgs boson or the scalar superpartners of the Standard Model) and a field with a shift symmetry. Shift-symmetry breaking effects like Yukawa couplings can provide the necessary spurion to make the diagrams nonzero, but we will generally have to go to two loops. As a result, the coefficients $a$ and $b$ may be of order $g^2/(16 \pi^2)$ for some marginal coupling $g$. Because this is a power divergence, a detailed number cannot be computed without a UV completion that regulates the loops. Because these are graviton loops, such a UV completion is not easy to come by, and will presumably require embedding the effective theory in a detailed string vacuum. Nonetheless, the general scaling argument should be trustworthy. From this we immediately extract one important lesson: {\em strongly suppressing scalar masses requires a cutoff well below the Planck scale}.

\begin{figure}[!h]
\begin{center}
\resizebox{0.5\textwidth}{!}{
\begin{tikzpicture}
\node at (-1,-0.25) {$\lambda$};
\node at (-0.2,0.75) {$\psi_\mu$};
\draw[fermion] (-1,0)--(1,0);
\draw[fermion] (3,0)--(1,0);
\draw[photon] (-1,0)--(3,0);
\draw[graviton] (1,0.76) circle (0.739);
\draw[fermion] (1,0.76) circle (0.739);
\node at (1,1.5) {\LARGE $\times$};
\node at (1,1.8) {$m_{3/2}$};
\node[blob] at (1,0) {};
\begin{scope}[shift={(4,0)}]
\node at (0,-0.25) {$\lambda$};
\node at (0.8,0.7) {$\psi_\mu$};
\node at (2,1.1) {\LARGE $\times$};
\node at (2,1.5) {$m_{3/2}$};
\draw[fermion] (0,0)--(1,0);
\draw[fermion] (4,0)--(3,0);
\draw[photon] (0,0)--(4,0);
\draw[graviton] (3,0.1) arc(0:180:1);
\draw[fermion] (3,0.1) arc(0:180:1);
\node[blob] at (1,0) {};
\node[blob] at (3,0) {};
\end{scope}
\end{tikzpicture} 
}
\end{center}
\caption{Gravitational loop corrections to gaugino masses. A gravitino mass insertion is required to break chiral and $R$ symmetries.} \label{fig:gauginogravcorr}
\end{figure}
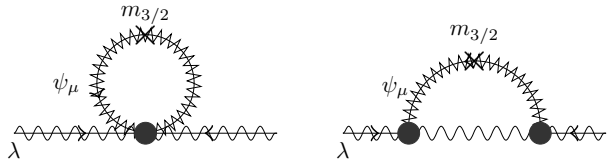
For gaugino masses, the estimate is even simpler, since we require an insertion of the gravitino mass to break chiral symmetry, as in Fig.~\ref{fig:gauginogravcorr}. Again, dimensional analysis provides the estimate of the divergent radiative correction to the gaugino mass
\beq
\delta_{\rm grav} m_\lambda \approx \frac{m_{3/2}}{16\pi^2 M_{\rm Pl}^2} \left(c \Lambda^2 + d m_{3/2}^2 \log \Lambda\right),
\eeq
where $c, d$ are again constant factors that we expect to be order-one or at least $g^2/(16 \pi^2)$. Again, {\em sequestering gravitino masses requires a low cutoff}.

These simple dimensional estimates provide the first indication that a consistent power-counting may exist in
\beq
\epsilon \equiv \frac{\Lambda}{4\pi M_{\rm Pl}}.
\eeq
If we are to avoid fine-tuning, we have the two inequalities:
\beq
m_{\rm scalar} & \simgt & \epsilon m_{3/2} \\
m_{\rm gaugino} & \simgt & \epsilon^2 m_{3/2}.
\eeq
How can we achieve a low cutoff and thus a small $\epsilon$? One approach is to UV complete four-dimensional gravity at a scale well below the 4d Planck scale. This can happen in the presence of extra dimensions bigger than Planck size, in which case the higher-dimensional theory has a smaller Planck scale, or if string states enter the calculation. (In fact, these two effects happen at similar scales in models with an order-one string coupling.) An interesting alternative is conformal sequestering~\cite{Luty:2001jh,Luty:2001zv,Schmaltz:2006qs}, which cuts off the loops in a different way: it postulates that the treatment of fields like $\phi$ and $\lambda$ as nearly-free propagating fields is incorrect due to the presence of strong dynamics and large anomalous dimensions. This approach to sequestering has received a great deal of attention from an effective field theory viewpoint. We will focus on the other case of low quantum gravity scales (large extra dimensions and light strings).

The arguments we have just given strengthen the case for the scenarios discussed in \S\ref{sec:dimred}: dimensional reduction of 5d supergravity or 10d Type IIB supergravity with large internal volumes will produce precisely the sort of hierarchies between the cutoff and the 4d Planck scale that are needed for consistently small radiative corrections.

\subsection{SUSY's ladder}
\label{sec:ladder}

Given our finding of no-scale structure from a large extra dimensional 
compactification and the dimensional analysis of  
gravitino loop corrections in  \S\ref{sec:gravitinoloop}, we see that the different 
scales may be separated naturally. The gravitino loops tell us the largest possible hierarchy 
among gauginos, scalars and gravitinos are
\begin{eqnarray}
   m_{\rm gaugino} \sim  \epsilon^2 m_{3/2} \ , \quad   
   m_{\rm scalar}  \sim  \epsilon m_{3/2} \ .
\end{eqnarray} 
Here we assume that loop corrections are the dominant contribution to the masses of sparticles, or at least are parametrically of the same order as the 
other ones (discounting non-parametric numerical factors like $1/(16\pi^2)$). The small parameter $\epsilon \equiv \frac{\Lambda} { 4 \pi M_{\rm Pl}}$, where $\Lambda$ is the Planck scale in 10d  (or 
the string scale). In 5d compactifications, it is unclear if we should take $\Lambda$ to be the KK scale (at which point we have to switch from a perturbative 4d theory to a perturbative 5d theory) or the 5d Planck scale (at which point local field theory breaks down entirely). In 10d compactifications, we will find that even when the overall volume is large there can be substructures within the manifold where the KK scale approaches the 10d Planck scale, so we expect that the 10d Planck scale is (for generic questions) a reasonable choice of cutoff. (Similar reasoning was used in \cite{Cicoli:2007xp}.) In the 5d case, for the moment we will take $\Lambda$ to be the 5d Planck scale, although we will not study a detailed explicit example and the KK scale may prove more appropriate in such a model. 

The dimensional reduction links the Planck scale in $D$-dimensions (which we will conflate with the string scale, assuming order-one coupling for the moment)
to the Planck scale in 4d, and meanwhile introduces the KK scale. 
Due to the large volume compactification, the string scale is much smaller than 
the Planck scale: 
\begin{equation}
   M_{\rm string}  \sim \frac{1}{ \sqrt{ \cal V}} M_{\rm Pl}  \simeq  \epsilon M_{\rm Pl}\ .
\end{equation}
Recall that ${\cal V}$ is the volume of the internal geometry in string units.
The next lower scale is the mass of Kaluza-Klein modes,
\begin{equation}
   M_{KK} \sim \frac{  M_{\rm string}  }{L} \sim  \frac{ M_{\rm string}}{ {\cal V}^{1/n} } \sim \frac{1 }
         { {\cal V}^{1/2+ 1/n} } M_{\rm Pl}  \sim \epsilon^{1+2/n} M_{\rm Pl}.
\end{equation}
In order to have a valid 4d effective supergravity theory, 
Although we need a model of moduli stabilization to fill 
in the details, we can estimate the mass of the gravitino
\begin{equation}
   m_{3/2} = \frac{1}{M_{\rm Pl}^2 }\langle e^{K/(2M_{\rm Pl}^2)} W \rangle 
         \sim \frac{M_{\rm Pl}} { \langle T + T^\dagger \rangle^{3/2} }   \ , 
\end{equation}
where the second equality uses the fact that no-scale structure has 
$K = -3 M_{\rm Pl}^2 \log ( T+ T^\dagger)$ and we take the natural assumption that $W$ takes values
near the Planck scale. Therefore, we have a relation between $M_{\rm Pl}$ and $m_{3/2}$,
\begin{eqnarray}  
   10d : m_{3/2}  \sim \epsilon^2 M_{\rm Pl} 
   \\
   5d : m_{3/2}  \sim \epsilon^3 M_{\rm Pl} 
\end{eqnarray}  
In both cases, the gravitino mass is less than or equal to the KK mass. These estimates allow us to complete the parametric estimates of the spectrum of SUSY's Ladder as presented in Figure \ref{fig:ladder}.

\section{Deviations from no-scale structure and moduli stabilization}
\label{sec:deviations}
In this section, we will add some sources of corrections to no-scale structure from the effective field theory point of view, 
which lead to deviations of $F_\Phi$ away from zero.  Having the corrections 
in effective field theory, we study the moduli stabilization including the large volume modulus and 
other moduli. The stabilization gives the moduli field F-terms and their masses. Based on these F-terms, we are able to 
study the soft SUSY breaking terms in the following section.  

\subsection{Effective field theory}

There are several possible sources of corrections to the leading-order no-scale 
structure, which include:
\begin{itemize}
\item {\em Instanton effects contributing $\int d^2 \theta~e^{-b {\bf T}}$ superpotential terms.} These arise, for example, from wrapping the worldvolumes of fields charged under the $p$-form in the higher-dimensional theory around the compactified dimensions.
\item {\em Additional moduli of the internal geometry.} Assumptions like ${\cal V} \propto ({\rm Re}~T)^{3/2}$ are generally not exact, as the internal geometry $h_{lm}(x,y)$ in general can give rise to multiple light scalar fields in the lower-dimensional theory. These also modify the effective K\"ahler potential in four dimensions.
\item {\em Higher-dimension operators in the higher-dimensional theory.} The supergravity action we started with includes the Einstein-Hilbert action at lowest order, but can include corrections involving higher powers of the Riemann curvature tensor. These give rise to modifications of the K\"ahler potential after reducing to four dimensions.
\item {\em Deviations from Ricci-flat internal geometries.} Fluxes, warping, and other effects can modify the internal geometry away from our assumption of Ricci-flatness, again modifying the four-dimensional effective theory. (Warping is potentially very interesting for phenomenology but we will have little to say about it here.)
\item {\em Uplifting.} In the end, we would like to obtain a 4d vacuum with nearly zero cosmological constant. This requires adding extra sources of SUSY breaking to lift the AdS vacuum we will find. We will not discuss details of the uplifting scenario in this paper, but it is important to check that it does not spoil no-scale structure (see \cite{Aparicio:2014wxa} for promising models).
\end{itemize}
All of these considerations lead us to expect that no-scale structure will be present in the four-dimensional theory only approximately. Deviations from no-scale structure can lead us in two different directions. The first, which has been widely pursued in the string theory literature, is to attempt to derive the structure of corrections from the top down. Some aspects of the literature on moduli stabilization remain controversial. A complementary approach is to work from the bottom up: rather than asking if a particular string theory construction gives rise to a controlled theory with stabilized moduli, we can ask whether an effective theory with approximate no-scale structure can be {\em internally} consistent. That is, is there a consistent power-counting for the suppressed effects such that corrections in the presence of higher-dimension operators and loops do not change the qualitative results? In the remainder of this section, we will take a largely top-down approach motivated by compactifying higher-dimensional supergravity. Later, in \S\ref{sec:eftoutlook}, we will take a preliminary look at loop corrections to the effective K\"ahler potential that allow us to ask about quantum consistency from the bottom-up.

First of all, the instanton effects are non-perturbative 
effects, where the instanton action is $\int d^2 \theta~e^{-b {\bf T}}$.
Considering only one volume modulus will lead us to the KKLT setup.
For two or more moduli fields, it is possible that the instanton action of small volume moduli is dominant over the action of the large volume modulus controlling no-scale structure; 
thus we will neglect the large volume modulus instantons. Starting from the volume in string units of ``swiss-cheese'' type with the big volume modulus ${\bf T}_b$ and the small volume modulus ${\bf T}_s$,
\begin{equation}
 {\cal V} = \left(T_b + T_b^\dagger\right)^{3/2} - 
         \left(T_s + T_s^\dagger\right)^{ 3/2} \ , 
\end{equation}
the Lagrangian having a small and large volume modulus is written as,  
\begin{equation}
   \mathcal{L} \simeq   - 3 \int d^4\theta  {\bf \Phi}^\dagger {\bf \Phi}
   M_*^2 \left[  \left( {\bf T}_b + {\bf T}_b^\dagger \right) - \frac{2}{3}   \frac{
         ( {\bf T}_s +  {\bf T}_s^\dagger)^{3/2}}
      { ( {\bf T}_b + {\bf T}_b^\dagger )^{1/2} } \right]
    + \left[ \int d^2\theta {\bf \Phi}^3 \left( W_0 + A e^{-a {\bf T}_s} \right)
   + {\rm h.c.} \right] \ .
\end{equation}

Secondly, higher order terms of the curvatures in the action will modify
the no-scale structure as well. We only consider the higher order terms in 10d, since in 5d 
the extra dimensional curvature is trivial (unless we have warped geometry, which would be interesting to explore in more detail). 
Since there are several different curvatures to consider, we use the notation $\tilde{R}$ for the curvature of $h_{lm}$ for the extra dimensions and ${R}_D$ for the curvature
in the original 10d theory. Note that $\tilde{R}$ does not include the volume modulus $L$. Without introducing ambiguity, $\tilde{R}$ and 
$R_D$ can be the Ricci or scalar curvatures or the Riemann tensor.  
We start with the dimensional reduction as in \S\ref{subsec:kk}. By only checking the kinetic terms of the volume modulus $L$, 
we will know the terms relevant to the corrections to K\"{a}hler potential.
Considering the Ricci-flat extra dimension, 
i.e. 
$\tilde{R}_{mn} =  \tilde{R} = 0 $,
the ${R}_D^2$ terms will
not introduce the kinetic term $\partial_\mu L \partial^\mu L$
in the 4d Lagrangian. Accordingly, it does not contribute to K\"{a}hler potential.
For ${R}_D^3$ terms, the leading order of $\partial_\mu (\log L )\partial^\mu ( \log L )$
is given by contracting three Riemann curvature tensors, one of which can be  
replaced by the kinetic term of $L$. It is straightforward to generalize the ${R}_D^3$ 
case to ${R}_D^{m+1}$, where $m \geq 2$. Each additional curvature tensor necessitates an additional inverse metric to raise an index when we form a 10d Lorentz invariant, and this brings along a factor of $L^{-2}$ after dimensional reduction. In the keinstein frame  
the action after compactification is written schematically as 
\begin{equation}
   \mathcal{L}_{\rm kein} \propto -\frac{1} { 16 \pi G_d}
      \int d^d x L^{ \alpha   -2 m }
      \left( \frac{\tilde{R}} { \tilde{M}_{\rm Pl}^2} \right)^m
    \left( \partial_\mu ( \log L)  \right)^2   \ ,
\end{equation}
where Lorentz contractions are implied. (In general there may be multiple terms with different contractions.)
$\tilde{M}_{pl}$ is the Planck scale in $10$d (or is the string scale---at the moment we are not being careful with factors of $g_s$). 
When $\tilde{R}$ and $\tilde{M}_{\rm Pl}$ are the same order, this term is only suppressed by $L^{\alpha - 2m}$.
In Type IIB supergravity the leading correction originates from ${R}_D^4$ operators,
since the ${R}_D^2$ and ${R}_D^3$ terms are incompatible with the $N=(2,0)$ supersymmetry of the 10d Lagrangian. (The specific coefficient of the ${R}_D^4$ terms is calculable when supergravity is UV completed to string theory \cite{Gross:1986iv}.) And after the compactification, the $\tilde{R}^3$ term coupled to 
$\partial_\mu (\log L )\partial^\mu ( \log  L )$ gives the corrections to K\"{a}hler potential. 
By removing the Ricci-flat assumption, the ${R}_D^2$ terms can modify the 
kinematic term of $\lambda$, and the correction is 
$\mathcal{L}_{\rm keinstein} \propto
-\frac{1} { 16 \pi G_d} \int d^d x L^{ \alpha - 2 }
\tilde{R} \frac{ \left( \partial_\mu L  \right)^2 } { L^2} $. Writing these corrections in superfields and recalling that in the 10d context we require $\alpha = 4$, the 
higher order terms of the curvatures modify the K\"{a}hler potential as follows:
\begin{equation}
   \mathcal{L}_{\rm higher-order} \simeq -  3 \int d\theta^4  {\bf \Phi}^\dagger {\bf \Phi}
  M_*^2 \xi  ( {\bf T}_b + {\bf T}_b^\dagger )^p 
\end{equation}
${R}_D^2$ terms lead to $p = 1/2$; ${R}_D^3$ terms lead to $p =0$~(logarithm); 
${R}_D^4$ terms lead to $p= -1/2$. In short, the $1/L^2$ factor coming from each additional Riemann tensor in 10d becomes a factor of $1/({\bf T}_b + {\bf T}_b^\dagger)^{1/2}$ in superspace.
Because ${R}_D^4$ is the leading nonzero correction allowed by the 10d supersymmetry, we take $p= -1/2$.

Lastly, in Type IIB supergravity we must also include an additional modulus arising from the 10d graviton multiplet, namely the dilaton-axion,
$S= e^{-\phi} + i C_0 $. Adding the above corrections and the modulus $S$,
the K\"{a}hler potential and superpotential are written as,
\begin{eqnarray}
   \mathcal{L} &\simeq& -  3\int d\theta^4  {\bf \Phi}^\dagger {\bf \Phi}
  M_*^2  \left( {\bf S} + {\bf S}^\dagger \right)^{1/3}
          \left[  \left( {\bf T}_b + {\bf T}_b^\dagger \right) - \frac{2}{3}   \frac{
         ( {\bf T}_s +  {\bf T}_s^\dagger)^{3/2}}
      { ( {\bf T}_b + {\bf T}_b^\dagger )^{1/2} } + \xi ({\bf S} + {\bf S}^\dagger)^{3/2}
         ( {\bf T}_b + {\bf T}_b^\dagger )^{-1/2}  \right]
      \nonumber \\
  &&
    + \int d\theta^2 {\bf \Phi}^3 \left( W_0 + A e^{-a {\bf T}_s}  + \frac{1}{2} W_s ( {\bf S} - S_0 )^2 \right)
   + {\rm h.c.}  \ .
\label{eq:lag_dilaton}
\end{eqnarray}    
The way that the dilaton couples to gravity is dictated by the symmetries of Type IIB supergravity. The leading order of the string coupling to 
$R_D$ and $R_D^4$ terms are the same (in string theory, this is the statement that both terms exist already at tree level---i.e.~that the leading $R_D^4$ term is an $\alpha'$ correction). The factor of $(S+ S^\dagger)^{3/2}$ results from the transformation from 10d string frame 
to 10d Einstein frame, $g_{M N}^{(\rm string)} \rightarrow e^{\frac{1}{2} \phi} g^{\rm (Einstein)}_{MN}$. The superpotential term $\frac{1}{2} W_s ( {\bf S} - S_0 )^2$ serves as a stand-in for the physics that stabilizes the dilaton in a supersymmetric manner, which can be accomplished via fluxes \cite{Giddings:2001yu}. We write a simple mass term because it leads to the right qualitative results. 
We can also consider string loop corrections as in \cite{Berg:2007wt,Cicoli:2007xp}, but we will defer preliminary comments on this until \S\ref{sec:eftoutlook} and leave a more detailed study for future work.

\subsection{Moduli stabilization in keinstein frame}
\label{sec:modulistabilize}

After having the Lagrangian derived from effective field theory, we can compute the F-terms and the potential of moduli 
in order to study moduli stabilization. In this section we will derive F-terms in KKLT and in Large Volume scenario, 
and estimate the mass of the moduli. The computation becomes much more clear by using the CDT gauge.

\subsubsection{KKLT}

We begin by considering the theory of a single modulus with no-scale K\"ahler term and KKLT-like superpotential~\cite{Luty:1999cz,Chacko:2000fn,Kachru:2003aw},
\beq
{\cal L} = - 3 \int d^4 \theta {\bf \Phi}^\dagger {\bf \Phi} M_*^2 \left({\bf T} + {\bf T}^\dagger\right) + \left[\int d^2 \theta {\bf \Phi}^3 \left(W_0 + A e^{-a {\bf T}} \right) + {\rm h.c.} \right].
\eeq
Reading off terms involving only homogeneous background values of the fields, this is:
\beq
- 3 M_*^2 \left(F_\Phi^\dagger F_\Phi \left(T + T^\dagger\right) + F_\Phi^\dagger \Phi F_T + \Phi^\dagger F_\Phi F_T^\dagger\right) + \left[3 \Phi^2 F_\Phi \left(W_0 + A e^{-a T} \right) - \Phi^3 a A e^{-a T} F_T + {\rm h.c.}\right].
\eeq
Varying with respect to the auxiliary fields leads to two equations of motion:
\beq
\frac{\delta}{\delta F_T^\dagger} & : & - 3 M_*^2 \Phi^\dagger F_\Phi - \Phi^{\dagger 3} a A e^{-a T^\dagger} = 0, \\
\frac{\delta}{\delta F_\Phi^\dagger} & : & - 3 M_*^2 \left(F_\Phi \left(T + T^\dagger\right) + \Phi F_T\right) + 3 \Phi^{\dagger 2} \left(W_0 + A e^{-a T^\dagger}\right) = 0.
\eeq
We solve the first of these equations for $F_\Phi$ and then use this to solve the second for $F_T$:
\beq
F_\Phi & = & -\Phi^{\dagger 2}\frac{a A}{3 M_*^2} e^{-a T^\dagger},\\
F_T & = & \frac{1}{3 M_*^2} \frac{\Phi^{\dagger}}{\Phi}\left[3\Phi^\dagger \left(W_0 + A e^{-a T^\dagger}\right) + a A e^{-a T^\dagger}\left(T + T^\dagger\right)\right].
\eeq
The minimization of potential with respect to $T$ gives us that $W = - \frac{1}{3} A e^{-a T^\dagger} 
\left[ 5+ a  ( T + T^\dagger) \right]$, leading to
$\frac{\langle F_\Phi \rangle}{ \langle \Phi  \rangle } \gg \frac{ \langle F_T \rangle } { \langle T + T^\dagger \rangle } $. It 
deviates from the result $F_\Phi = 0 $ from pure no-scale structure too much, and loses many of the desired phenomenological consequences. This simple example illustrates that no-scale structure may be easily broken if the modulus ${\bf T}$ is strongly stabilized by the superpotential. 

\subsubsection{Large Volume Scenario}

Now, let us consider the Large Volume Scenario (LVS)~\cite{Balasubramanian:2005zx,Conlon:2005ki}. In contrast to KKLT,
it has small deviation from the no-scale structure. Essentially, the key point is that the superpotential terms involving the large modulus ${\bf T}_b$ are so tiny that they can be neglected, so the stabilization of ${\bf T}_b$ is via the K\"ahler potential. 
We start from the Lagrangian derived from effective field theory having large and small volume moduli, a dilaton-axion modulus, and a
higher dimension operator from $R^4$ as in eq.~(\ref{eq:lag_dilaton}), which we repeat here for convenience:
\begin{eqnarray}
   \mathcal{L} &\simeq& -  3\int d\theta^4  {\bf \Phi}^\dagger {\bf \Phi}
  M_*^2  \left( {\bf S} + {\bf S}^\dagger \right)^{1/3}
          \left[  \left( {\bf T}_b + {\bf T}_b^\dagger \right) - \frac{2}{3}   \frac{
         ( {\bf T}_s +  {\bf T}_s^\dagger)^{3/2}}
      { ( {\bf T}_b + {\bf T}_b^\dagger )^{1/2} } + \xi ({\bf S} + {\bf S}^\dagger)^{3/2}
         ( {\bf T}_b + {\bf T}_b^\dagger )^{-1/2}  \right]
      \nonumber \\
  &&
    + \int d\theta^2 {\bf \Phi}^3 \left( W_0 + A e^{-a {\bf T}_s}  + \frac{1}{2} W_s ( {\bf S} - S_0 )^2 \right)
   + {\rm h.c.}  \ . \nonumber
\label{eq:lag_dilaton_repeated}
\end{eqnarray}    
We work under the assumption that $  \frac{ \langle F_{T_b} \rangle } {  \langle T_b + T_b^\dagger \rangle } \gg 
   \frac{ \langle F_S \rangle } {  \langle S + S^\dagger \rangle } ,
   \frac{ \langle F_\Phi \rangle } { \langle \Phi  \rangle} , $ which will be verified to be self-consistent afterwards.
We read off the leading auxiliary terms (neglecting fermion fields), 
\begin{eqnarray}
  F_{T_b}^\dagger  & : &  - 3M_*^2 \frac{F_{T_b}^\dagger}  {T_b+ T_b^\dagger} 
            \left[  \frac{3}{4}\frac{ F_{T_b} } { T_b + T_b^\dagger} 
            \frac{ - \frac{2}{3} \left( T_s + T_s^\dagger \right)^{3/2}   + \xi \left( S + S^\dagger \right)^{3/2} }
         { \left( T_b + T_b^\dagger  \right)^{1/2} }  
         + \frac{1}{2}\frac{ F_{T_s} } { T_s + T_s^\dagger}    \frac{  \left( T_s + T_s^\dagger \right)^{3/2} }
         { \left( T_b + T_b^\dagger  \right)^{1/2} }  
         \right]
   \label{eq:LVS_Ftb}
   \\
  F_S^\dagger  & : &  - M_*^2 \frac{F_S^\dagger}{S+ S^\dagger} 
         \frac{ F_{T_b} } { T_b + T_b^\dagger}
             \Phi  \Phi^\dagger   ( S+ S^\dagger )^{1/3} ( T+ T^\dagger) 
         +  F_S^\dagger {\Phi^\dagger}^3 M_s \left( \langle S^\dagger \rangle - S_0^\dagger \right)
   \label{eq:LVS_FS}
   \\
  F_\Phi^\dagger  & : &  - 3 M_*^2 \frac{F_\Phi^\dagger}{\Phi^\dagger} 
               \frac{ F_{T_b} } { T_b + T_b^\dagger}
             \Phi  \Phi^\dagger   ( S+ S^\dagger )^{1/3} ( T+ T^\dagger) + 3  \frac{F_\Phi^\dagger}{\Phi^\dagger} 
            {\Phi^\dagger}^3 W 
   \label{eq:LVS_Fphi}
   \\
   F_{T_s}^\dagger &  : &  - 3 M_*^2 \frac{F_{T_s}^\dagger}{T_s+ T_s^\dagger} 
            \left( 
            \frac{1}{2}\frac{ F_{T_b} } { T_b + T_b^\dagger}
            - \frac{1}{2}\frac{ F_{T_s} } { T_s + T_s^\dagger}
         \right)
             \Phi  \Phi^\dagger   ( S+ S^\dagger )^{1/3} \frac{ \left( T_s + T_s^\dagger \right)^{3/2}} 
         { \left( T_b + T_b^\dagger  \right)^{1/2} }  - {F_{T_s}^{\dagger} }{\Phi^\dagger}^3 a A e^{- a T_s^{\dagger} }    
   \label{eq:LVS_Fts}
      \ , 
\end{eqnarray}
where $W  = W_0 + A e^{-a T_s } + \frac{1}{2} W_s ( S - S_0)^2 $.
We have two ways to compute $F_{T_b}$ by varying the auxiliary fields of $F_S^\dagger$, $F_\Phi^\dagger$  
in eq.~(\ref{eq:LVS_FS},\ref{eq:LVS_Fphi}) respectively,
\begin{equation}
   \frac{\langle F_{T_b} \rangle} { \langle T_b + T_b^\dagger \rangle}  \simeq \frac{ { \langle \Phi^\dagger \rangle}^2 \langle W \rangle } { M_*^2 \langle \Phi \rangle  \langle S + S^\dagger \rangle^{1/3}  { \langle T_b + T_b^\dagger \rangle} }
     \simeq \frac{ {\langle \Phi^\dagger \rangle}^2 M_S \left( \langle S^\dagger \rangle - S_0^\dagger \right)  \langle S + S^\dagger \rangle } {  M_*^2 \langle \Phi \rangle  \langle S + S^\dagger \rangle^{1/3}  { \langle T_b + T_b^\dagger \rangle}  }
   \ ,
   \label{eq:Ftb}
\end{equation}
which leads to the relation of $ \langle W_0 \rangle  \simeq  
   M_S \left( \langle S^\dagger \rangle - S_0^\dagger \right)  \langle S + S^\dagger \rangle $. 
This can be derived by minimizing the potential in the $T_b$ direction, or the leading order can be given by the supersymmetric minimum
$D_S W = 0$. 
From the eq.~(\ref{eq:LVS_Ftb}), we can see that 
\begin{eqnarray}
      \frac{ \langle F_{T_b} \rangle } {  \langle T_b + T_b^\dagger \rangle }   \sim \mathcal{O} (1) \times  
   \frac{\langle F_{T_s} \rangle}{  \langle T_s + T_s^\dagger \rangle } 
   \sim m_{3/2}
\end{eqnarray}
The other F-terms may be found by considering the next-to-leading order term in a large-volume expansion of the equation of motion:
\begin{eqnarray}
   \frac{\langle F_S \rangle}{  \langle S + S^\dagger \rangle } 
      & \simeq  & 
         \frac{  -  {\langle \Phi^\dagger \rangle}^2  a A e^{ - a \langle T_s^\dagger\rangle}  } {  M_*^2 \langle \Phi \rangle \langle S + S^\dagger \rangle^{1/3} }
            \frac{ \langle T_s + T_s^\dagger \rangle} {  \langle T_b + T_b^\dagger \rangle}
         \sim \frac{ m_{3/2} } {\cal V}
      \\
      \frac{\langle F_\Phi \rangle } { \langle \Phi  \rangle} &\simeq &
         \frac{  -  3 {\langle \Phi^\dagger \rangle}^2 \xi  \langle S + S^\dagger \rangle^{3/2}  \langle W \rangle   } 
              {  4 M_*^2 \langle \Phi \rangle \langle S + S^\dagger \rangle^{1/3}   \langle T_b + T_b^\dagger \rangle^{5/2}     }
         \sim \frac{ m_{3/2} } {\cal V}
\end{eqnarray}

Having calculated all the F-terms, we can compute the mass of moduli. For the large volume modulus, the first term of the K\"ahler
potential in eq.~(\ref{eq:lag_dilaton}) in CDT gauge gives us its leading kinetic term, and the second term is the leading mass term of (the real part of) $T_b$.
Normalizing the field $T_b$ canonically leads to  
\begin{equation}
   m_{T_b} \sim  \frac{ m_{3/2}} { {\cal V}^{1/2} } \ .
\end{equation}
For the small volume modulus $T_s$, the second term of the K\"ahler potential in eq~(\ref{eq:lag_dilaton}) gives both the kinetic
term and mass, hence 
\begin{equation}
   m_{T_s} \sim  m_{3/2}   \ .
\end{equation}
It turns out that $S$ has larger supersymmetric mass than its soft mass, which is consistent with integrating out $\bf S$ supersymmetrically as in 
in \S\ref{subsec:integrateout}. Its mass can be derived from eq.~(\ref{eq:lag_dilaton}) as well: 
\begin{equation}
   m_{S} \sim \frac{ \sqrt{W_S M_{\rm Pl} }}  { {\cal V}} \ , 
\end{equation}
on the same order as $m_{3/2}$ if $W_S \sim M_{\rm Pl}$. In all of these cases, the scaling of the mass in terms of powers of volume is straightforwardly read off from the superspace Lagrangian, without apparent cancelations.

\section{Soft SUSY breaking from superspace}
\label{sec:susybreaking}

Computations of the soft SUSY breaking effective Lagrangian in various incarnations of the LVS have been discussed in refs.~\cite{Conlon:2005ki,Blumenhagen:2009gk,Conlon:2006us,Conlon:2006wz,Conlon:2007dw,Cicoli:2012aq,Higaki:2012ar,Aparicio:2014wxa}, following the lines of general earlier work on supersymmetry breaking in supergravity theories~\cite{deCarlos:1992pd,Kaplunovsky:1993rd,Brignole:1993dj,Kaplunovsky:1994fg,Randall:1998uk,Giudice:1998xp}. Many of the computations presented in the literature involve nontrivial cancelations between different terms, often compensating terms proportional to $m_{3/2}$ with other terms related to the $F$ components of the moduli fields. In \S\ref{sec:purenoscale}, we saw that at leading order these somewhat mysterious cancelations are absent when working in superspace, and originate from the fact that $F_\Phi = 0$ in the pure no-scale limit. In this section, we will look at corrections beyond the pure no-scale limit and see that the suppression of soft terms is easily read off from the Lagrangian and is compatible with the estimates based on loop calculations in \S\ref{sec:hierarchies}.

\subsection{Soft scalar masses}

We saw in \S\ref{subsec:sequesteredchiral} that in the pure no-scale limit soft masses are absent for sequestered chiral superfields. Away from the pure no-scale limit, $F_\phi \neq 0$ and we may also have new couplings of the chiral superfield to the modulus $\bf T_b$. Suppose the kinetic term for the chiral superfields is
\beq
{\cal L}_{\rm kin} = \int d^4 \theta\, {\bf \Phi}^\dagger {\bf \Phi} \left(1 + \frac{c_Q}{\left({\bf T_b} + {\bf T_b}^\dagger\right) ^{3/2}}\right) {\bf Q}^\dagger {\bf Q}  \ .  \label{eq:chiralkineticansatz}
\eeq

Here we have added a volume-suppressed term proportional to a coefficient $c_Q$ (which may in general depend on complex structure moduli, but for the moment we take to be a constant). 
The reason to add the volume-suppressed term is that this correction is consistent with the effective field theory perspective, 
which is estimated by gravitino loops in \S\ref{sec:gravitinoloop}.
Because $F_{T_b}$ is the dominant $F$-term in the theory, let us first consider the terms proportional to $F_{T_b}^\dagger F_{T_b}$. These arise from taking two derivatives:
\beq
{\cal L}_{\rm kin} \supset \Phi^\dagger \Phi \frac{15 c_Q}{4 \left(T_b + T_b^\dagger\right)^{7/2}} \left|F_{T_b}\right|^2 Q^\dagger Q.
\eeq
Now, the factor of $\Phi^\dagger \Phi$ appears in the leading kinetic term and hence serves to canonically normalize the scalar field. Taking derivatives brought in two new factors of $(T_b+T_b^\dagger)^{-1}$, but we recall that $m_{3/2} \approx \left<F_{T_b}/(T_b + T_b^\dagger)\right>$, so the role of these new factors is to translate the factors of $F_{T_b}$ into factors of $m_{3/2}$. Finally, we observe that the remaining powers of $T_b$ in the denominator are approximately just the volume of the internal geometry in string units. As a result, we have that
\beq
{\cal L}_{\rm kin} \supset \frac{15 c_Q}{4 {\cal V}} m_{3/2}^2 Q^{c\dagger} Q^c.
\eeq
Recalling that $\epsilon = 1/\sqrt{\cal V}$, we see that this is consistent with the scaling
\beq
m_{\rm scalar} \sim \epsilon m_{3/2}
\eeq
that we estimated from loops and dimensional analysis in \S\ref{sec:hierarchies}. There are additional terms $\sim F_\Phi^\dagger F_T + {\rm h.c.}$ and $\sim F_\Phi^2$, but they are subdominant.

Let us compare to the results of \cite{Aparicio:2014wxa}. They studied a K\"ahler potential of the form
\beq
K = -2 M_{\rm Pl}^2 \log({\cal V} + {\hat \xi}/2) - \frac{f_\alpha(U,S)}{{\cal V}^{2/3}} \left(1 - c_s \frac{\hat \xi}{\cal V}\right) Q^\dagger Q + \cdots.
\eeq
To compare to our ansatz (\ref{eq:chiralkineticansatz}) we must evaluate ${\bf \Omega} = -3M_{\rm Pl}^2 \exp(-{\bf K}/(3 M_{\rm Pl}^2))$. Expanding in the limit of large $\cal V$, we find that $\bf \Omega$ matches our ansatz if we rescale $\bf Q$ by a constant factor $\propto \left<f_\alpha\right>^{1/2}$ and make the choice
\beq
c_Q = \frac{1}{3} {\hat \xi} \left(c_s - \frac{1}{3}\right).
\eeq 
This explains why \cite{Aparicio:2014wxa} found that the scalar masses are suppressed at the special value $c_s = 1/3$: from the superspace point of view, this is the case $c_Q = 0$ when the volume-suppressed part of the $\bf Q$ kinetic term is simply absent. This is yet another example of how working in superspace can make the outcome of calculations more transparent. The case $c_Q = 0$ was referred to as the ``ultralocal limit'' in \cite{Aparicio:2014wxa} and further studied in more detail in \cite{Aparicio:2015sda}. It has the potential to produce {\em both} scalar and gaugino masses $\sim m_{3/2}/{\cal V}$, producing CMSSM-like phenomenology in a setting where the gravitino problem is decoupled. From our point of view, however, $c_Q = 0$ looks unnatural; the loop estimates in \S\ref{sec:hierarchies} suggest that it will not hold, barring UV physics that would effectively regulate the loop in a way that seems magical from the low-energy EFT point of view. On the other hand, $c_Q \ll 1$ might already be interesting, and could lead to a distorted SUSY's ladder where the scalar rung is a bit lower than expected. It would be interesting to pursue a more detailed understanding of the reasonable size of the coefficient $c_Q$ in the future.

\subsection{Gaugino masses}

Gaugino masses originate from the holomorphic gauge kinetic function $f_a$,
\be
\int d^2\theta \frac{1}{4} {\bf f}_a \mathbfcal{W}^{a\alpha} \mathbfcal{W}^a_\alpha + {\rm h.c.},
\ee
where the lowest component of $f_a$ controls the gauge coupling and theta angle: $\left<f_a\right> = \frac{1}{g_a^2} - \frac{i}{8\pi}\Theta_a$. Clearly, if $f_a$ contains a linear term in the moduli $T_b$ or $T_s$, we will obtain gaugino masses of order $m_{3/2}$. On the other hand, suppose that $f_a$ is controlled by the dilaton:
\be
{\bf f}_a = \delta_a {\bf S}.
\ee
Then we obtain gaugino masses
\be
m_{\lambda_a} = \frac{1}{2} \delta_a \left<\frac{F_S}{S}\right> \sim \frac{m_{3/2}}{\cal V}.
\ee
This is consistent with our estimate in \S\ref{sec:gravitinoloop} of the smallest possible gaugino-to-gravitino mass ratio that is not destabilized by loops. 

We also expect anomaly mediation to generate contributions of order $m_{\lambda_a} \sim (\alpha_a/\pi) \left<F_\Phi/\Phi\right>$, but because $\left<F_\Phi/\Phi\right> \sim m_{3/2}/{\cal V}$ these are a loop factor smaller than the tree-level dilaton contributions.

\subsection{A terms}

Consider a superpotential interaction like the top quark Yukawa coupling,
\be
\int d^2 \theta {\bf \Phi}^3 y_t {\bf H_u} {\bf Q_3} {\bf {\bar u}_3} + {\rm h.c.}
\ee
Recall that the leading kinetic terms for each of these fields will also involve the conformal compensator through a factor ${\bf \Phi}^\dagger {\bf \Phi}$, so the canonically normalized fields are rescaled, e.g.
\be
{\bf H_u}^c = \left<\Phi\right> {\bf H_u},
\ee
and we can write the Yukawa interaction as
\be
\int d^2 \theta \left(\frac{\bf \Phi}{\left<\Phi\right>}\right)^3 y_t {\bf H_u}^c {\bf Q_3}^c {\bf {\bar u}_3}^c + {\rm h.c.}
\ee
This shows that the Yukawa coupling $y_t$ can be an order-one number that does not scale with any power of the internal volume. From this we also read off that there will be an $A$-term
\be
A_t = \left<\frac{F_\Phi}{\Phi}\right> y_t \sim \frac{y_t m_{3/2}}{\cal V}.
\ee
Thus we expect to find $A$-terms on order of the gaugino masses times Yukawa couplings. Compared to the larger scalar masses of order $m_{3/2}/\sqrt{\cal V}$, these will have little dynamical role to play.

\subsection{$\boldsymbol{\mu}$ and b terms}

\subsubsection{Phenomenological requirements}

So far we have found that large-volume SUSY breaking can produce the hierarchies of SUSY's Ladder that we discussed in \S\ref{sec:ladder}: scalars at $m_{3/2}/\sqrt{\cal V}$ and gauginos at $m_{3/2}/{\cal V}$. To obtain realistic phenomenology, we must be able to arrange for one light Higgs doublet to play the role of the Standard Model Higgs. This places requirements on the values of $\mu$ and $b_\mu$ that we want to obtain. Recall that the mass matrix of the two Higgs doublets is
\be
{\cal M}_{\rm Higgs}^2 = \begin{pmatrix} \left|\mu\right|^2 + m_{H_u}^2 & b_\mu \\ b_\mu & \left|\mu\right|^2 + m_{H_d}^2 \end{pmatrix},
\ee
and the requirement of a light Standard Model-like doublet imposes
\be
\det {\cal M}_{\rm Higgs}^2 \ll \left({\rm tr} {\cal M}_{\rm Higgs}^2 \right)^2. \label{eq:getalighthiggs}
\ee
If $\left|\mu\right|^2 \ll \left|m_{H_u}^2\right|, \left|m_{H_d}^2\right|$, this tells us that we would like $b_\mu$ to be approximately the geometric mean of the soft masses squared. On the other hand, if we take $\mu$ to be of the same order as the scalar soft masses, there is a significant threshold correction to the gaugino masses \cite{Pierce:1996zz,Gherghetta:1999sw}:
\be
\delta M_{1,2} \sim \frac{\alpha_{1,2}}{2\pi} \mu \frac{b_\mu}{\mu^2 - m_A^2} \log\frac{\mu^2}{m_A^2}.
\ee
We cannot suppress this correction by taking $b_\mu$ small because then the requirement (\ref{eq:getalighthiggs}) leads to large $\tan \beta$ and we would predict a too-large Higgs mass given our heavy scalars. Thus taking $\mu \sim 1000~{\rm TeV}$ is in some tension with our desire to have gauginos near the weak scale for interesting phenomenology (including the possibility of SUSY dark matter). As a result, we hope to obtain hierarchies $\mu \ll m_{3/2}/\sqrt{\cal V}$ and $b_\mu \sim (m_{3/2}/\sqrt{\cal V})^2$.

\subsubsection{K\"ahler contributions}

Consider contributions to $\mu$ and $b_\mu$ from a holomorphic kinetic term
\be
\int d^4\theta\, {\bf \Phi}^\dagger {\bf \Phi} {\bf H_u} {\bf H_d} \left(c_{H,0} + \frac{c_{H,1}}{\left({\bf T}_b + {\bf T}_b^\dagger\right)^{3/2}} + \ldots \right) + {\rm h.c.}, \label{eq:kineticmubmu}
\ee
where the omitted terms correspond to further volume suppression. The factors $c_{H,0}$ and $c_{H,1}$ may depend on moduli but are assumed to be independent of ${\bf T}_b$ and ${\bf S}$ so that they do not contain large $F$-terms.

From this expression we can immediately read off a $\mu$-term
\be
\mu = c_{H,0} \left<\frac{F_\Phi^\dagger}{\Phi^\dagger}\right> - \frac{3 c_{H,1}}{2 {\cal V}}\left<\frac{F_{T_b}}{T_b}\right> \sim \frac{m_{3/2}}{\cal V},
\ee
which is on the same order as the gaugino masses and hence does not lead to problematic threshold corrections. If the $c_H$ coefficients depend on $S$ there may be additional contributions, but they are still of order $m_{3/2}/{\cal V}$.

We also read off the leading contribution to $b_\mu$,
\be
b_\mu = \frac{15 c_{H,1}}{4{\cal V}} \left<\frac{F_{T_b}}{T_b}\right>^2 \sim \frac{m_{3/2}^2}{\cal V}.
\ee
This is on the same order as the soft scalar masses, as is necessary to realize the condition (\ref{eq:getalighthiggs}) for obtaining a light Higgs boson. A shift symmetry in the Higgs kinetic terms could even guarantee a vanishing determinant at leading order by relating $c_{H,1}$ to the coefficients $c_{H_u}$ and $c_{H_d}$ in the ${\bf H_u}^\dagger {\bf H_u}$ and ${\bf H_d}^\dagger {\bf H_d}$ terms \cite{Hebecker:2012qp}, although the Higgs couplings explicitly break such a symmetry and we will still need to fine-tune $b_\mu$ in the end for realistic electroweak symmetry breaking.

\subsubsection{Superpotential contributions}

So far, working in superspace has allowed us to read off that the leading contributions to many SUSY-breaking terms is precisely of the order that we would like to see from a phenomenological viewpoint. However, there are some dangerous effects that could spoil this picture. In particular, superpotential contributions to $\mu$ and $b_\mu$ may be problematic. These can arise from a ``de-sequestering'' term like \cite{Berg:2010ha, Berg:2012aq, Aparicio:2014wxa}
\be
\int d^2 \theta\, {\bf \Phi}^3 W_H e^{-a {\bf T}_s}{\bf H_u} {\bf H_d} + {\rm h.c.} =  \int d^2 \theta\, \frac{{\bf \Phi}^3}{\left<\Phi\right>^2} W_H e^{-a {\bf T}_s}{\bf H_u}^c {\bf H_d}^c + {\rm h.c.}\label{eq:desequester}
\ee
If $W_H \sim M_{\rm Pl}$, then after canonically normalizing the Higgs fields we read off a contribution
\be
\mu_W = \left<\Phi\right> W_H \left<e^{-a T_s}\right> \sim \frac{1}{\left<T_b + T_b^\dagger\right>^{1/2}} M_{\rm Pl} \frac{1}{\cal V} \sim \frac{M_{\rm Pl}}{{\cal V}^{4/3}} \sim \frac{m_{3/2}}{{\cal V}^{1/3}}.
\ee
This is larger even than the soft scalar masses, and will lead to large threshold corrections to electroweak gaugino masses. As a result, we must demand that terms of the form (\ref{eq:desequester}) are absent. As emphasized in \cite{Aparicio:2014wxa}, if we replace the coefficient $a$ in the exponent with a coefficient $a' > 5a/3$ the term becomes safely small. A similar statement holds for the $b_\mu$ term which gets a contribution
\be
(b_\mu)_W \sim \mu_W \left<F_{T_s}\right> \sim \frac{m_{3/2}^2}{{\cal V}^{1/3}},
\ee
again undesirably large, but which becomes $\simlt m_{3/2}^2/{\cal V}$ if $a' > 5a/3$.

This ``de-sequestering'' problem implies that the familiar $\mu$-problem of supersymmetric model-building acquires new aspects in the context of extra-dimensional, sequestered model-building. What is encouraging is that the kinetic terms alone lead to unproblematic answers whenever there is a good inverse volume expansion as in (\ref{eq:kineticmubmu}). Thus, our problem is merely to explain the absence or suppression of superpotential terms like (\ref{eq:desequester}), which may follow from geometric properties of the compactification in particular UV models \cite{Cicoli:2012vw,Cicoli:2013cha}.

\section{Discussion}
\label{sec:discussion}

\subsection{Phenomenology of SUSY's Ladder}
\label{sec:splitversions}

The observation of the Higgs boson at 125 GeV restricts the range of scalar masses that make sense in split SUSY \cite{Giudice:2011cg,Bagnaschi:2014rsa,Vega:2015fna}. 
The heaviest admissible scalar masses are near $10^8~{\rm GeV}$ and require pushing $\tan \beta$ very close to 1 \cite{Bagnaschi:2014rsa}, which is unexpected given the tendency of RG running to split $m_{H_u}^2$ and $m_{H_d}^2$. (The scalar masses can be pushed slightly heavier given a large higgsino mass \cite{ArkaniHamed:2012gw}, though this in turn generates a large threshold correction to gaugino masses.) More plausible parameter regimes include scalars at 20 TeV with large $\tan \beta$ or scalars at 1000 TeV with $\tan \beta \approx 2$. The former regime has received intensive attention in a variety of ``mini-split'' scenarios where gaugino masses are a loop factor below scalar masses \cite{Giudice:1998xp,Wells:2003tf,Feldman:2011ud,Kane:2011kj,Arvanitaki:2012ps,ArkaniHamed:2012gw}. It seems fair to say that it has been widely viewed as the most well-motivated variation of split SUSY. It offers the tantalizing prospect of scalars just barely light enough that some may be discovered at a future 100 TeV collider \cite{Cohen:2014hxa, Ellis:2015xba}.

SUSY's Ladder has scalars around the 1000 TeV scale, which can explain the Higgs mass at relatively small values of $\tan \beta$. It is known that universal scalar masses $m_{H_u}^2 = m_{H_d}^2 = m_{Q_3}^2 = \ldots$ at the GUT scale lead to the correct Higgs mass when scalar masses are at about a PeV and $\tan \beta \approx 2$ \cite{Bagnaschi:2014rsa}. This is encouraging for our scenario if we expect universal scalar masses at the GUT scale. There are two aspects to this question: the first is whether the leading couplings of MSSM scalars to the modulus ${\bf T}_b$ are universal and the second is which UV scale is relevant for imposing boundary conditions on the RGEs. These questions are beyond the scope of this paper. However, Conlon and Palti \cite{Conlon:2009xf,Conlon:2009kt,Conlon:2009qa} have studied threshold corrections in string theory and concluded that large threshold effects push the effective universality scale to ${\cal V}^{1/6} M_{\rm string}$, which is close to the standard GUT scale in the SUSY's Ladder scenario. This is an encouraging hint that the Higgs mass of 125 GeV may emerge in a natural way from 10d no-scale compactifications. (Though, as always in split SUSY, with a tuning of $b_\mu$ to achieve a light Standard Model-like doublet.)

Our results motivate a closer look at the variation of split SUSY with scalars at 1000 TeV, which has received less attention, except when viewed as the upper end of mini-split when the bottom of the gaugino spectrum is at multiple TeV as required for thermal wino dark matter \cite{Wells:2004di}. Probing 1000 TeV scalars experimentally is an interesting challenge. Flavor and CP provide indirect probes \cite{Altmannshofer:2013lfa} and the 1000 TeV scale could be interesting from the viewpoint of explaining the SM flavor structure \cite{Baumgart:2014jya}. The gluino lifetime is in the hundred micron range, on the edge of accessibility with current detector technology \cite{Arvanitaki:2012ps,ArkaniHamed:2012gw}. It would be interesting to explore other possibilities, like whether a 100 TeV collider could measure the gluino pair production cross section accurately enough to measure the interference term arising from $t$-channel exchange of 1000 TeV squarks, or whether precision measurements of gaugino and higgsino decay branching fractions could contain a sufficient amount of information to indirectly probe the scalar spectrum. Of course, our starting point was the gravitino problem, and dark matter physics and cosmology can also probe this scenario. Inflationary phenomenology perhaps offers the best hope for direct access to some aspects of the moduli physics.

\subsection{Building an effective field theory}
\label{sec:eftoutlook}

In this paper we have taken some steps in the direction of building a convincing effective field theory of no-scale structure. However, there is more to do.  The dimensional analysis arguments of \S\ref{sec:hierarchies} provide some indication that the hierarchical SUSY's Ladder spectrum is radiatively stable. However, it would be useful to compute loop corrections directly in the context of the full theory (including both Standard Model fields and moduli fields and their couplings to each other). Certain aspects of these loop effects have been estimated in \cite{Berg:2007wt,Cicoli:2007xp,Burgess:2010sy}, which found that an ``extended no-scale structure'' ensures that the leading-order results are not wildly altered by loops. These results relied on analogy to computable string theory loops (in toroidal orientifolds) and on Coleman-Weinberg potential calculations with an appropriate UV cutoff.

Ideally, we would like to have a supersymmetric formalism for estimating the size of higher-dimension operators directly in superspace in the conformal compensator formalism we are working with. For now we will settle for making some comments based on the one-loop effective K\"ahler potential for a nonrenormalizable theory \cite{Brignole:2000kg}. This contains several terms, but we will focus on the quadratically divergent contribution
\be
\delta K = \frac{\Lambda^2}{16\pi^2} \log \det {\hat K},
\ee
where ${\hat K}_{i^\dagger j} = \frac{\partial^2 K}{\partial Q^{i^\dagger} \partial Q^j}$ is the matrix of second derivatives of the K\"ahler potential.  Unfortunately, several things are unclear about this expression, including the appropriate choice of cutoff $\Lambda$ and the appropriate scale to make the argument of the logarithm dimensionless. Let us forge ahead and try to make conservative choices.

Consider a no-scale K\"ahler potential with sequestered chiral matter, $K = -3 M_{\rm Pl}^2 \log(T + T^\dagger - Q^\dagger Q)$. In this case the matrix of second derivatives of the K\"ahler potential is
\be
{\hat K} = 3 M_{\rm Pl}^2 \begin{pmatrix} \frac{1}{(T + T^\dagger - Q^\dagger Q)^2} & \frac{-Q^\dagger}{(T + T^\dagger - Q^\dagger Q)^2} \\ \frac{-Q}{(T + T^\dagger - Q^\dagger Q)^2} & \frac{T + T^\dagger}{(T + T^\dagger - Q^\dagger Q)^2} \end{pmatrix}.
\ee
The tricky part of the calculation is that $\Lambda$ is viewed as a constant in the original discussion of the one-loop effective K\"ahler potential, but reasonable choices in our context include the string scale or KK scale which, in Planck units, depend on the expectation value $\left<T\right>$. We will assume that $\Lambda$ is in fact a function of the {\em field} $T$. The {\em largest} reasonable cutoff we can take is $M_{\rm string} = M_{\rm Pl}/{\cal V}^{1/2} = M_{\rm Pl}/(T+T^\dagger)^{3/4}$. Taking $\Lambda = \beta M_{\rm string}$, with $\beta$ an ${\cal O}(1)$ number reflecting our uncertainty about ultraviolet physics, we have
\be
\delta K = -\frac{3\beta^2 M_{\rm Pl}^2}{16\pi^2 (T+T^\dagger)^{3/2}} \left[\log(T + T^\dagger - Q^\dagger Q) + \gamma\right],
\ee 
where $\gamma$ is a constant that depends on the scale of the argument of the logarithm. The correction $\delta K$ is volume-suppressed compared to the leading-order K\"ahler potential. This can be rephrased as a superspace kinetic function (expanding around large $T$):
\be
{\bf \Omega} = {\bf T} + {\bf T}^\dagger - {\bf Q}^\dagger {\bf Q} - \frac{\beta^2}{16\pi^2} \frac{\gamma + \log({\bf T} + {\bf T}^\dagger)}{({\bf T} + {\bf T}^\dagger)^{1/2}} + \frac{\beta^2}{16\pi^2} \frac{\gamma +1 + \log({\bf T} + {\bf T}^\dagger)}{\left({\bf T} + {\bf T}^\dagger\right)^{3/2}}  {\bf Q}^\dagger {\bf Q} + \ldots
\ee
Apart from factors of $\log({\bf T} + {\bf T}^\dagger)$ in coefficients, we see that the leading correction to the pure ${\bf T}$-dependence of ${\bf \Omega}$ is a $({\bf T} + {\bf T}^\dagger)^{-1/2}$ term, like the one we included from $R^4$ operators in the 10d theory in equation (\ref{eq:lag_dilaton}). The leading correction to the ${\bf Q}$ kinetic term is supressed by $({\bf T} + {\bf T}^\dagger)^{-3/2}$, as we assumed in (\ref{eq:chiralkineticansatz}). Thus, taking the largest possible cutoff in the computation of the effective K\"ahler potential gives further justification for the choices we have made above. Loop corrections do not generate dangerous operators like ${\bf Q}^\dagger {\bf Q}/({\bf T} + {\bf T}^\dagger)^{1/2}$, for instance, which would completely spoil the phenomenology. The underlying reason for this is easy to see: the dangerous terms are quadratic divergences proportional to $\Lambda^2 \sim M_{\rm Pl}^2/{\cal V}$, so all of the corrections are suppressed by the volume rather than some other power of the length scale of internal dimensions. A related estimate of the shift in moduli masses due to Planck-suppressed couplings to gauge fields was performed in \cite{Krall:2014dba}, which again found sufficient volume suppression to preserve the phenomenology.

As emphasized in \cite{Cicoli:2007xp}, we might expect loop effects on small cycles to be sensitive to the string-scale radius of those cycles while loop effects from KK modes of the large volume might be cut off at the lower scale $M_{\rm KK} \sim M_{\rm string}/{\cal V}^{1/6}$. This can only decrease the importance of loops, so our estimates have been pessimistic. On the other hand,  \cite{Berg:2007wt,Cicoli:2007xp} discuss a possible $1/(T+T^\dagger)$ term in the effective K\"ahler potential, which superficially appears dangerous but does not change the potential. In ${\bf \Omega}$, such a term is simply a constant, so again superspace clarifies why it is harmless. However, our effective K\"ahler potential estimate has so far turned up no indication that such a term exists.

Although our first look at loop corrections has been encouraging, there is much more to understand. If different cutoffs $\Lambda$ appear in different parts of the calculation, the standard effective K\"ahler potential formalism will not give correct answers. The field-dependence of the cutoff, the uncertain argument of the logarithm, and the fact that we are working in a gravitational theory rather than global supersymmetry all suggest the need for a more powerful formalism for computing loops. 

\subsection{Further explorations}
\label{sec:further}

We have focused on the phenomenology of large volume compactifications of Type IIB supergravity. Our general argument in \S\ref{sec:dimred} also suggested the possibility that 5d compactifications could lead to no-scale structure. Here there is some ambiguity about whether the cutoff scale in our estimates should be viewed as the KK scale or the 5d Planck scale, which would lead to rather different phenomenology. UV completions of the 5d scenario could naturally exist in heterotic M-theory. It would be very interesting to explore these possibilities in more detail in the future. We have also omitted the uplifting sector, which must be part of any complete model that is capable of canceling the cosmological constant. 

More generally, we hope that our results can help to bridge the gap between activity in Large Volume superstring compactifications and in phenomenological studies of supersymmetry. The phenomenology community has to date shown limited interest in the results of string phenomenology. One reason for this is justified concern about whether the approximations being used will stand up to corrections: the idea of a model with $m_{\rm gaugino} \ll m_{3/2}$ understandably makes theorists worry. Our results suggest that these concerns can be answered in an effective field theory with a controlled power-counting, the key feature being a cutoff that is parametrically below the Planck scale (which happens naturally in certain extra-dimensional theories). Although many of our results have previously been derived in the string phenomenology literature, we hope that by presenting them in a different formalism where the scaling of various terms with the small parameter $\epsilon = 1/\sqrt{\cal V}$ is completely manifest, we have made it more clear that apparently mysterious cancelations are in fact automatic outcomes. No uncanny ``string magic'' is at work, just effective field theory in a gravitational context. De-sequestering terms in the superpotential are a real concern, but K\"ahler corrections seem to be under control. By using the formalism and notation of superspace, we hope to have clarified the underlying physics and made it more accessible to phenomenologists wary of delving into the literature on string theory.

As the LHC continues to test supersymmetry, completely natural models may begin to fall by the wayside. It could be that the notorious cosmological problems induced by gravitinos and moduli provide a partial explanation for why we do not live in a completely natural universe. If we allow a single moderate tuning in the Higgs boson mass, the SUSY's Ladder scenario can naturally avoid such cosmological problems while explaining the existence of large hierarchies in nature. The first experimental signal would likely be the discovery of a gluino. We eagerly await further data from the 13 TeV LHC.

\section*{Acknowledgments}
We thank Marcus Berg, Michele Cicoli, Ben Heidenreich, M.C.~David Marsh, Brent Nelson, Enrico Pajer,  Fernando Quevedo, and Jesse Thaler for useful discussions. MR would especially like to thank Michele Cicoli for emphasizing the possibility of suppressed volume modulus couplings to dark matter in the Large Volume Scenario in discussions at the MCTP Hidden Dark Matter Conference, which stimulated MR's renewed interest in this topic. MR thanks the organizers and participants of that conference and of String Pheno Cosmo 2015 for providing congenial settings to discuss the interplay between phenomenology and string theory. The work of MR is supported in part by the NSF Grant PHY-1415548 and the NASA Grant 14-ATP14-0018. The work of WX is support by the U.S. Department of Energy under grant Contract Numbers DE-SC00012567 and DE-SC0013999.

\appendix
\section{The heterotic string, M-theory, and no-scale structure}
\label{app:heterotic}

We have argued in \S\ref{sec:dimred} that no-scale structure suggests a preferred role for Type IIB superstrings, because of the existence of a four-form gauge field which can give rise to the imaginary part of the chiral superfield that serves as a K\"ahler modulus. However, the first discussion of no-scale structure in string theory that we are aware of dates all the way back to Witten's 1985 work on dimensional reduction of heterotic $E_8 \times E_8$ superstrings~\cite{Witten:1985xb}. Witten found the K\"ahler potential
\beq
K/M_{\rm Pl}^2 = -3 \log(T + T^\dagger) - \log(S+S^\dagger)
\eeq
where ${\rm Re}~T = e^{\sigma} \phi^{3/4}$ and ${\rm Re}~S = e^{3\sigma} \phi^{-3/4}$, where $e^{\sigma/2}$ is the length scale of the six internal Calabi-Yau dimensions (analogous to our $L(x)$) and $\phi$ is the dilaton or string coupling constant. From this we see that the field $T$ has the desired no-scale kinetic term. If the superfield $S$ were to acquire a large supersymmetric mass, we can consistently set it equal to its VEV and perhaps obtain no-scale phenomenology (as in \S\ref{subsec:integrateout}). What lesson should we draw from this? Our discussion in \S\ref{sec:dimred} assumed that we were interested in a single field describing an isotropic length scale for the internal dimensions. Theories with two or more fields, such as the dilaton and the length scale of the internal dimensions, potentially give rise to a wider variety of ways to realize no-scale structure.

However, this example in fact {\em is} a disguised form of one of the examples we derived above.  The strong coupling limit of the heterotic $E_8 \times E_8$ superstring is heterotic M-theory~\cite{Horava:1995qa,Witten:1995em,Witten:1996mz,Horava:1996ma}, which at low energies is 11-dimensional supergravity compactified on $S^1/\mathbb{Z}_2$. We can study the 11-dimensional theory reduced to four dimensions on a Calabi-Yau times an interval, or alternatively we can study it reduced on the Calabi-Yau to a five-dimensional orbifold theory~\cite{Lukas:1997fg,Lukas:1998yy,Lukas:1998tt}. Suppose that we begin with the following ansatz for an 11-dimensional metric, choosing our notation to resemble~\cite{Lukas:1997fg}:
\beq
ds^2 = {\bar g}_{\mu \nu}(x) dx^\mu dx^\nu + e^{2 a(x)} h_{lm}(y) dy^l dy^m + e^{2 c(x)} dx^{11} dx^{11}.
\eeq
If we work in a regime where the six dimensions described by the $y^m$ coordinates are much smaller than the remaining five dimensions, we can dimensionally reduce to obtain a 5d action
\beq
S_{5d} = -\frac{1}{16 \pi G_{11}} \int d^5 x \sqrt{-{\tilde g}} V_6 e^{6 a(x)} \left({\tilde {\cal R}}_5 + 30{\tilde g}^{M N} \partial_M a \partial_N a\right),
\eeq
with ${\tilde g}_{MN} dx^M dx^N = {\bar g}_{\mu \nu}(x) dx^\mu dx^\nu + e^{2 c(x)} dx^{11} dx^{11}$. We can rewrite this in the five-dimensional Einstein frame via the Weyl transformation ${\tilde g}_{MN} = e^{-4 a(x)} g_{MN}$, i.e.
\beq
g_{MN} dx^M dx^N & = & g_{\mu \nu}(x) dx^\mu dx^\nu + e^{2 {\hat c}(x)} dx^{11} dx^{11},~{\rm where} \nonumber \\
g_{\mu \nu} = e^{-4a} {\bar g}_{\mu \nu} &,& {\hat c}(x) = c(x) + 2 a(x).
\eeq
It is precisely the field $e^{{\hat c}(x)}$ which becomes ${\rm Re}~T(x)$ when we repackage all of the fields into chiral supermultiplets. In other words, the somewhat mysterious combination $e^{\sigma} \phi^{3/4}$ from Witten's heterotic string reduction is precisely the same thing as the length scale of the $S^1/{\mathbb Z}_2$ direction in heterotic M-theory, as measured in the 5d Einstein frame after integrating out the Kaluza-Klein modes associated with the Calabi-Yau dimensions. The superfield $S$ describes the remaining dimensions.

This shows that the appearance of no-scale structure from a peculiar combination of two fields in the heterotic string can be understood due to the continuous connection of that theory with a simpler five-dimensional effective field theory to which our earlier discussion applies. The remaining ingredient is that the imaginary part of the superfield $T$ is associated with a 1-form gauge field in five dimensions, which arises from integrating the underlying 3-form gauge field of M-theory over a two-dimensional cycle in the Calabi-Yau.

\bibliography{ref}
\bibliographystyle{utphys}
\end{document}